\documentclass[%
preprint,
showpacs,preprintnumbers,
nofootinbib,
amsmath,amssymb,
aps,
]{revtex4-1}

\usepackage{graphicx}
\usepackage{dcolumn}
\usepackage{bm}
\usepackage{hyperref}
\usepackage{amssymb,amsmath,amsfonts,amsxtra,amsthm}
\usepackage{graphicx,color}
\usepackage{dcolumn}
\usepackage{array}
\usepackage{tabularx}
\usepackage{enumerate}
\usepackage{hhline}
\usepackage{subfigure}

\begin{document}
	
	
	\title{Peak statistics for the primordial black hole abundance
	}
	
	\author{Yi-Peng Wu$^{a}$}
	\email{ywu@lpthe.jussieu.fr}
	\affiliation{$^{a}$Laboratoire de Physique Th\'{e}orique et Hautes Energies (LPTHE), \\
		UMR 7589 CNRS \& Sorbonne Universit\'{e}, 4 Place Jussieu, F-75252, Paris, France}

\date{\today}

\begin{abstract}
	The primordial black hole (PBH) abundance evaluated by the conventional Press-Schechter (PS) probability distribution is shown to be equivalent to the high-peak limit of a special point-like peak statistics via unphysical dimensionality reduction of the Bardeen-Bond-Kaiser-Szalay (BBKS) theory.
	The fact that PBHs are formed at high peak values $\nu_c \gg 1$ leads to a systematic bias proportional to $\nu_c^3$ between the predictions of the PS method and the physical BBKS peak theory in a general three-dimensional spatial configuration.
	As long as realistic PBHs are collapsed from three-dimensional density peaks in space, the systematic bias led by $\nu_c^3$ implies a significant underestimation of the PBH abundance reported by the PS method.
	For the inflationary spectrum in the narrow-spike class, the underestimation in the extended mass functions is further enlarged by at least a factor of $10^{2.5}$ in all mass range, indicating a severer constraint to models in the favor of considering PBHs as all dark matter in a certain mass range.

\end{abstract}

\pacs{04.50.Kd, 98.80.Jk}
\maketitle

\section{Introduction}
\label{sec:intro}

Primordial black holes (PBHs) \cite{Zeldoovich:1967,Hawking:1971ei,Carr:1974nx,Carr:2009jm,Khlopov:2008qy} are dark matter that can span a wide range of mass scales. The question whether PBHs occupy a significant fraction of the total dark matter density becomes a revival topic since the indirect detections of stellar mass black holes by LIGO and Virgo \cite{Carr:2016drx}. While BHs around $10 - 10^2 M_\odot$ are still possible to provide up to $10$ percent in the dark matter density ratio $f = \rho_{\rm PBH}/\rho_{\rm DM}$, the updating constraints indicate that the window for realizing $f = 1$ by monochromatic mass PBHs could be opened around the sub-lunar mass range ($ 10^{-16}-10^{-11} M_\odot$) \cite{Dasgupta:2019cae}, see also \cite{Niikura:2017zjd,Katz:2018zrn,Bai:2018bej,Jung:2019fcs,Laha:2019ssq,Arbey:2019vqx,Boudaud:2018hqb,Montero-Camacho:2019jte,Belotsky:2018wph}.

A narrow but spiky mass distribution is favorable for the purpose of having PBH dark matter in a certain mass range \cite{Bartolo:2018evs,Cai:2018dig,Saito:2008jc,Wang:2019kaf,Inomata:2017okj,Inomata:2017vxo,Frampton:2010sw,Garcia-Bellido:2016dkw,Garcia-Bellido:2017aan}. However, even if one starts with a delta-function like spectrum of the curvature perturbation $\zeta$ generated from inflation, the final density $f = f(M)$ at matter-radiation equality inevitably spreads out a distribution many orders in the BH mass $M$ (see \cite{Wang:2019kaf} as an example), essentially due to the effect of critical collapse \cite{Yokoyama:1998xd,Niemeyer:1997mt}.
There are several uncertainties coming from the non-Gaussianity of $\zeta$ \cite{DeLuca:2019qsy,Atal:2018neu,Saito:2008em,PinaAvelino:2005rm,Franciolini:2018vbk,Tada:2015noa,Young:2015cyn,Byrnes:2012yx,Bullock:1996at,Yokoyama:1998pt,Garcia-Bellido:2016dkw}, the choice of the smoothing window function for the density perturbation $\Delta$ \cite{Ando:2018qdb}, the non-linearity in transferring $\zeta$ to $\Delta$ \cite{Young:2019yug,Kawasaki:2019mbl,Kalaja:2019uju}, the initial profile of $\Delta$ \cite{Kalaja:2019uju,Yoo:2018esr} and the threshold $\Delta_c$ of the gravitational collapse \cite{Carr:1975qj,Niemeyer:1999ak,Musco:2004ak,Musco:2008hv,Musco:2012au,Nakama:2013ica,Harada:2013epa,Musco:2018rwt,Escriva:2019nsa,Escriva:2019phb}, that can all affect the the final mass distribution from a given inflationary power spectrum $\mathcal{P}_\zeta$. For preciseness, our choice of the density contrast $\Delta$ is the same as defined in \cite{Young:2014ana}.

Recently, many efforts have been made to improve the conventional computation for the PBH mass function based on the Press-Schechter (PS) approach \cite{Press:1973iz} or the Bardeen-Bond-Kaiser-Szalay (BBKS) statistics of density peaks \cite{Bardeen:1985tr}.
 To verify a more accurate probability distribution of peaks valid for PBH formation, it is argued that one should take into account the spatial correlation with the threshold of gravitational collapse $\Delta_c$ \cite{Germani:2019zez}, or the smoothing scale correlation with the peak value $\nu = \Delta/\sigma_\Delta$ \cite{Young:2020xmk,Germani:2019zez,Suyama:2019npc}, where $\sigma_\Delta$ is the variance for $\Delta$. The conditional statistics of peaks at the maximum along the smoothing scale could help to avoid the so-called cloud-in-cloud (or BH-in-BH) problem \cite{Suyama:2019npc}, yet such an issue might have only negligible corrections to the mass function from broad inflationary spectrum \cite{DeLuca:2020ioi}.


Constraining inflationary models via the PBH mass function requires careful treatment on all of the uncertainties (see \cite{Kalaja:2019uju} for the first attempt with the monochromatic mass assumption). To meet a desirable BH abundance for dark matter, one expects that the amplitude of $\mathcal{P}_\zeta$ in a narrow-spike shape must be much larger than that in a board shape. This fact can be attributed to the shape dependence of the peak value $\nu_c \equiv \Delta_c/\sigma_\Delta$ \cite{Germani:2018jgr}.
A huge discrepancy for the resulting PBH abundance was reported in \cite{Germani:2018jgr}, between predictions from the BBKS method and the PS method.
Ambiguities arise, however, as pioneering studies \cite{Green:2004wb,Young:2014ana} comparing the two methods at a fixed $\nu_c$ found a systematic discrepancy within the uncertainties of finding the exact threshold $\Delta_c$. It should be notice that Refs. \cite{Green:2004wb,Young:2014ana} focused on blue-tilted spectra, which can be generated, for example, by curvaton scenarios \cite{Kawasaki:2013xsa} (two-field inflation models). On the other hand, models of single-field inflation for PBH formation typically create red spectra \cite{Saito:2008em,Yokoyama:1998pt,Germani:2017bcs,Garcia-Bellido:2017mdw,Cicoli:2018asa}, where most of the spectral shapes can be well-approximated by the broken power-law templates \cite{Atal:2018neu}. 

The fact that peaks must be maxima among the extrema of the density perturbation imposes natural constraints to the probability distribution, which sources the essential difference between the BBKS method and the PS method, where the latter treats $\Delta$ as an independent random field to the spatial configuration. Such an extremum-to-maximum (E-to-M) condition, promotes the BBKS method to become a higher-dimensional (multivariate) statistics involved with a scalar random field $\Delta$ and its first and second spatial derivatives. In this work, we outline the role of the E-to-M condition in the PBH mass function.

In order to get a comprehensive understanding on the systematic difference between the PS and BBKS methods, we formulate a special statistics of  point-like peaks with respect to the E-to-M condition, where each peak has zero dimension in space and has exact spherical symmetry. 
The point-like peak theory not only reduces the number of random variables of the BBKS method but also reproduces the correct (yet unphysical) spatial dimensionality of the PS statistics.
We treat, $\nu_c$, the peak value of collapse, as an independent entity to the probability distribution of peaks, due to the fact that the density variance $\sigma_\Delta$ is solely determined by the input $\mathcal{P}_\zeta$ with a smoothing strategy, whereas the threshold $\Delta_c$ must rely on numerical simulations based on a selected density profile (despite that the density profile is correlated with the input spectrum \cite{Bardeen:1985tr,Germani:2019zez}).

The statistics of point-like peaks derived in this work is convenient for extracting out the effect of the E-to-M condition on the PS method and the impact of dimensional reduction to the BBKS method.
We employ useful templates for the inflationary power spectrum, including the broken power-law spectrum for single-field inflation \cite{Atal:2018neu}, to compute the PBH density at each Hubble scale and the extended mass function seen at matter-radiation equility. Our templates cover the spectral shape from narrow to broad and from red to blue tilted in the momentum space.
In this paper, we report the systematic bias in the PBH number density and the shape dependence of such a systematic bias from various models of the inflationary spectrum.
With the help of the special point-like peak theory, we conclude the (in)accuracy of the PBH abundance estimated via the PS method.

\section{Templates for inflationary spectrum}

In this section we prepare templates of the inflationary power spectrum that will be applied to compute the PBH abundance via methods with different statistical bias.
As a first example, we consider a spectral template that summarized well-studied models of PBH formation in the framework of single-field inflation. The template is a special case of the so-called broken power-law type \cite{Clesse:2018ogk,Atal:2018neu} of the form 
\begin{equation}\label{template_power_spectrum}
\mathcal{P}_\zeta(k) = \left\lbrace \begin{array}{cr}
A_\zeta \left(k/k_0\right)^{n-1} &, \quad k\geq k_0, \\
0 &,\quad k < k_0,
\end{array}\right.
\end{equation}
where $A_\zeta$ and $n$ are parameters of the spectral amplitude and index, respectively. Current models of single-field inflation are realized in the range of $-2 < n < 1$, see \cite{Atal:2018neu}. Note that $\zeta$ is the gauge invariant variable which coincides with the comoving curvature perturbation (defined from the spatial part of metric) on uniform density hypersurfaces. 

To remove the effect of (very broad) superhorizon fluctuations, it is pointed out in \cite{Young:2014ana} that the density contrast, denoted by $\Delta$, is the appropriate quantity for the discussion of BH formation in both theories. 
The spectrum of $\Delta$, $\mathcal{P}_\Delta$, connects to the power spectrum of curvature perturbation through the relation
\begin{align}
\mathcal{P}_\Delta (k,t) = \frac{4(1+w)^2}{(5+3w)^2}\left(\frac{k}{aH}\right)^4 \mathcal{P}_\zeta(k).
\end{align}
The variance of the density contrast is computed by
\begin{align}
\sigma^2_\Delta(R,t) = \int_{0}^{\infty} W^2(kR) \mathcal{P}_\Delta(k,t) d\ln k,
\end{align}
where we have applied a window function $W(kR)$ with a smoothing scale $R$ to avoid the divergence in the large-$k$ limit. 
For convenience, we adopt the Gaussian window function $W(kR) = \exp[-k^2R^2/2]$, and we shall fix the smoothing scale with the comoving horizon as $R = 1/(aH)$. Therefore the variance can be computed from a given spectrum $\mathcal{P}_\zeta$ as
\begin{align}\label{variance_gaussian_window}
\sigma^2_\Delta(R) = \frac{4(1+w)^2}{(5+3w)^2}\int_{0}^{\infty} e^{-k^2R^2} \left(kR\right)^4\mathcal{P}_\zeta(k) d\ln k.
\end{align} 
The $i$-th spectral moment, $\sigma_i = \sigma_i(R)$, is smoothed by the Gaussian window function according to
\begin{align}
\sigma_i^2 (R)= \int_{0}^{\infty} k^{2i} W^2(kR) \mathcal{P}_\Delta(k, R) d\ln k,
\end{align}
where $\sigma_0 = \sigma_\Delta$. We focus on the statistical properties of density peaks in radiation domination with $w =1/3$. 

\begin{figure}
	\begin{center}
		\includegraphics[width=70mm]{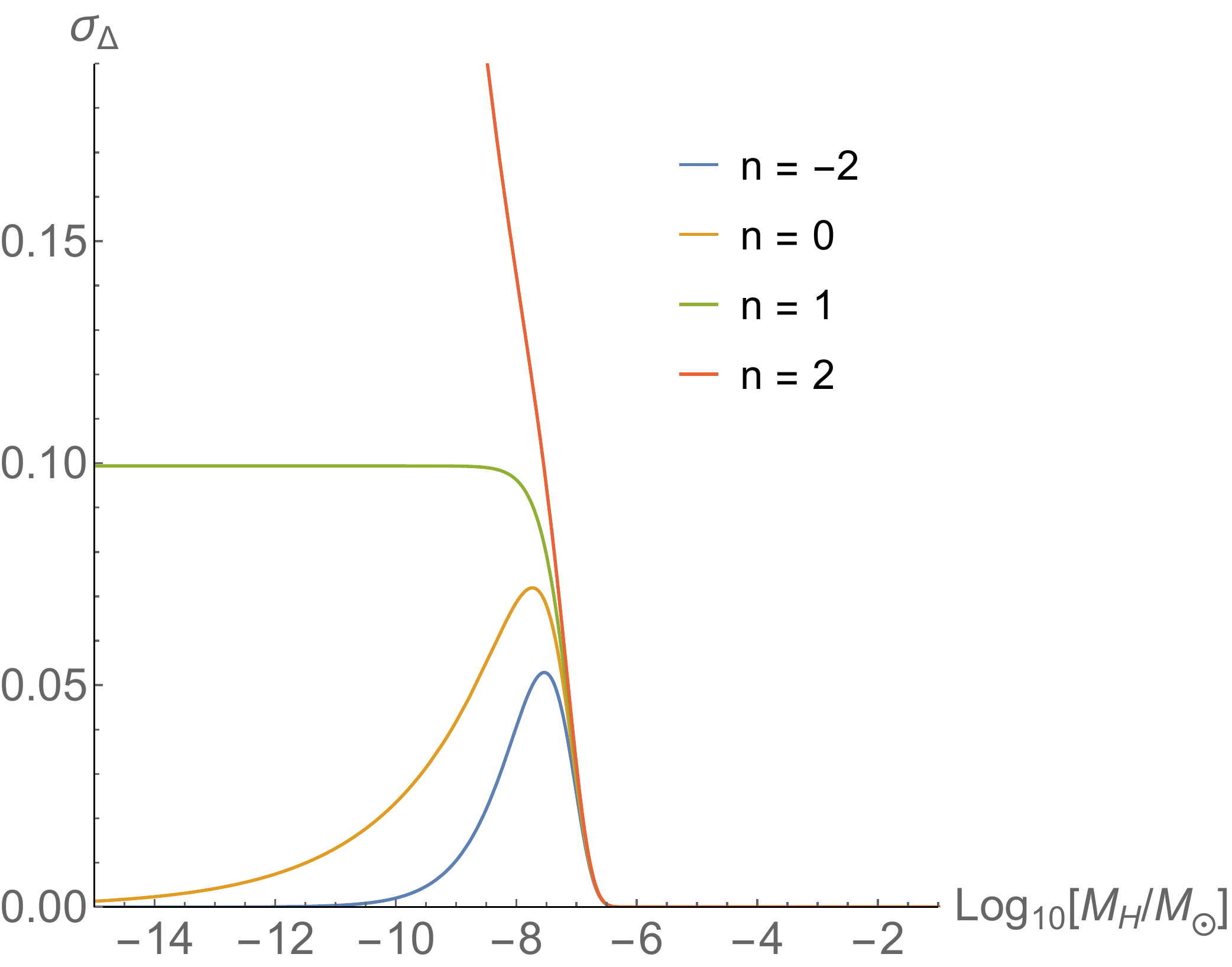}
		\hfill
		\includegraphics[width=70mm]{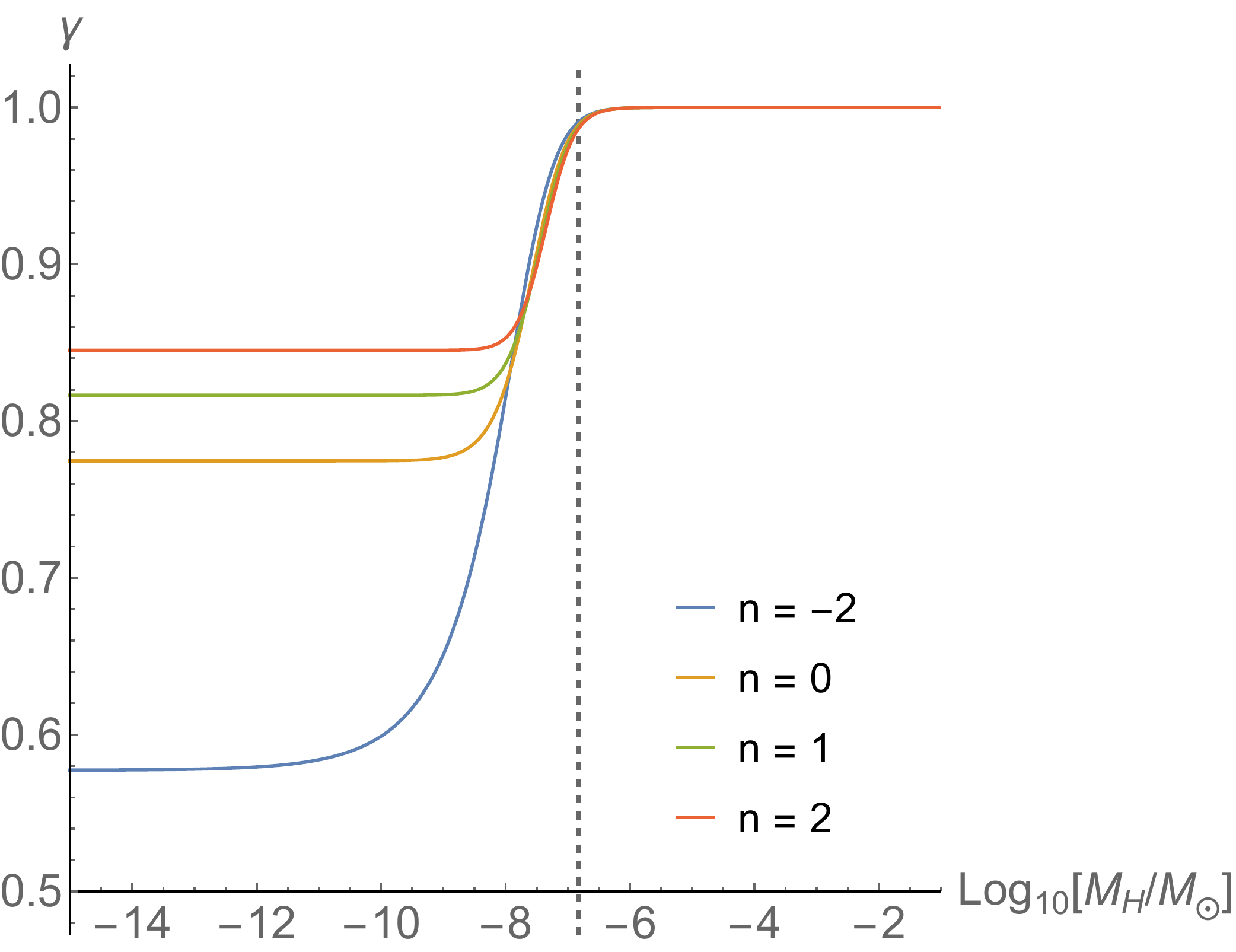}
	\end{center}
	\caption{
		\label{fig_BPL}The variance $\sigma_\Delta (M_H)$ (left panel) and the $\gamma$ factor (right panel) of the broken power-law template with the pivot scale $k_0$ corresponding to a horizon mass $ M_{H0} = 1.5\times 10^{-7} M_\odot$ (dashed line).  
	}
\end{figure}

One can obtain the variance of the template \eqref{template_power_spectrum} accoriding to \eqref{variance_gaussian_window} with $w =1/3$ as
\begin{align}\label{variance_single-field}
\sigma^2_\Delta(R) = \frac{4}{81}\frac{A_\zeta}{2}(k_0R)^{1-n}\, \Gamma\left(\frac{n+3}{2}, k_0^2R^2\right),
\end{align} 
where $\Gamma(a, z)$ is the incomplete gamma function.
Here we have applied the useful formula for the integration as
\begin{align}
\int_{r_0}^{\infty}e^{-r^2}r^m \left(\frac{r}{r_0}\right)^{n-1}dr = \frac{r_0^{1-n}}{2} \Gamma\left(\frac{m+n}{2}, r_0^2\right),
\end{align}
where the change of variables $r = k R$ and $r_0 = k_0 R$ are used.
Similarly, the first spectral moment is found to be
\begin{align}\label{first_moment_single-field}
\sigma^2_1(R) = \frac{4}{81}\frac{A_\zeta}{2 R^2}(k_0R)^{1-n} \Gamma\left(\frac{n+5}{2}, k_0^2R^2\right).
\end{align} 
Note that $\Gamma(a, z) \rightarrow \Gamma(a)$ as $z \rightarrow 0$ and thus the results of the power-law spectrum investigated in \cite{Young:2014ana} can be reproduced by taking $k_0 R \rightarrow 0$ to the above results. 
We show examples of $\sigma_\Delta(R)$ computed from the broken power-law template in Figure~\ref{fig_BPL} 
with different choices of $n$, where $R = R(M_H)$ is given by \eqref{horizon_mass to smoothing}. A pivot scale $k_0$ can convert to a pivot horizon mass $M_{H0}$ in the Solar unit according to
\begin{align}
\frac{k_0}{k_{eq}} = 5.29 \times 10^{8}\left(\frac{M_\odot}{M_{H0}}\right)^{1/2}\left(\frac{g_{\ast eq}}{g_{\ast 0}}\right)^{1/6}.
\end{align}
The choice with $n = 1$ stands for a step spectrum and in the cases with $n > 1$ the spectrum is blue-tilted for $k > k_0$. In this work, we use $g_{\ast 0} = 106.75$ with $ M_{H0} = 1.5\times 10^{-7} M_\odot$, which corresponds to the horizon mass at temperature $T \simeq 300$ GeV. 


The broken power-law spectrum \eqref{template_power_spectrum} is ill defined for $n \geq 1$ as one has to put cutoff in the large $k$ limit by hand \cite{Young:2014ana,Byrnes:2018clq}.
This motivates us to consider an improved template for the blue spectrum in the form of trapezoidal shape as
\begin{equation}\label{template_top_hat}
\mathcal{P}_\zeta(k) = \left\lbrace \begin{array}{cr}
A_\zeta (k/k_{\rm min})^{n-1}, & \quad k_{\rm max}\geq k \geq k_{\rm min}, \\
0 . &\quad \mathrm{otherwise},
\end{array}\right.
\end{equation}
where we focus on $n \geq 1$ for this template. The common top-hat model \cite{Germani:2018jgr,Saito:2009jt} can be recovered at $n = 1$.
The spectral moments of this model can be computed readily via a subtraction of the broken power-law results at different pivot scales. For example, the variance reads
\begin{align}
\sigma_\Delta^2(R) =& \frac{4}{81}\frac{A_\zeta}{2} (k_{\rm min }R)^{1-n} \\\nonumber
&\times \left[\Gamma\left(\frac{3+n}{2}, k_{\rm min}^2R^2\right) - \Gamma\left(\frac{3+n}{2}, k_{\rm max}^2R^2\right)\right],
\end{align}
where the amplitude at $k_{\rm max}$ is $A_\zeta(k_{\rm max}/k_{\rm min})^{n-1}$.


\begin{figure}
	\begin{center}
		\includegraphics[width=70mm]{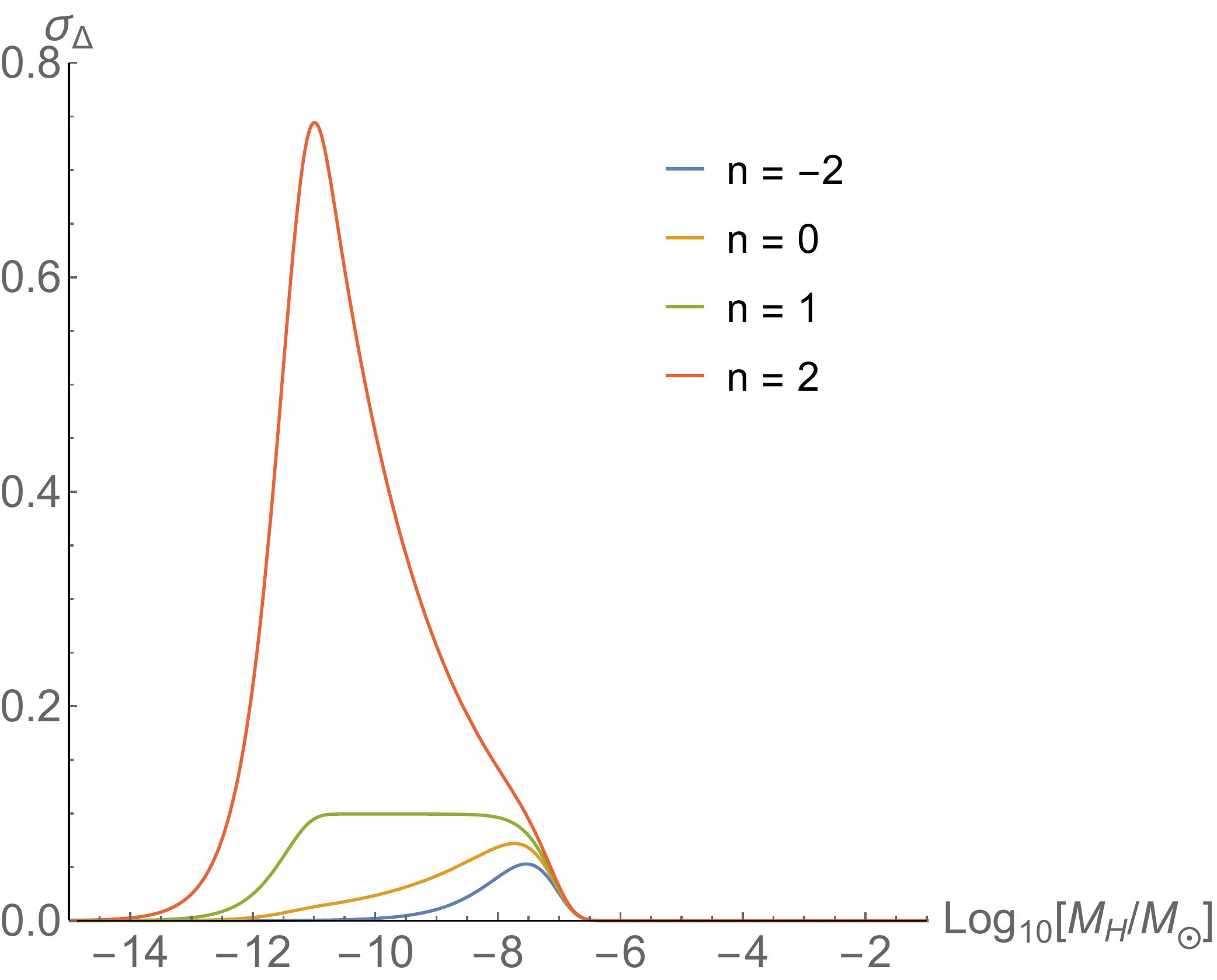}
		\hfill
		\includegraphics[width=70mm]{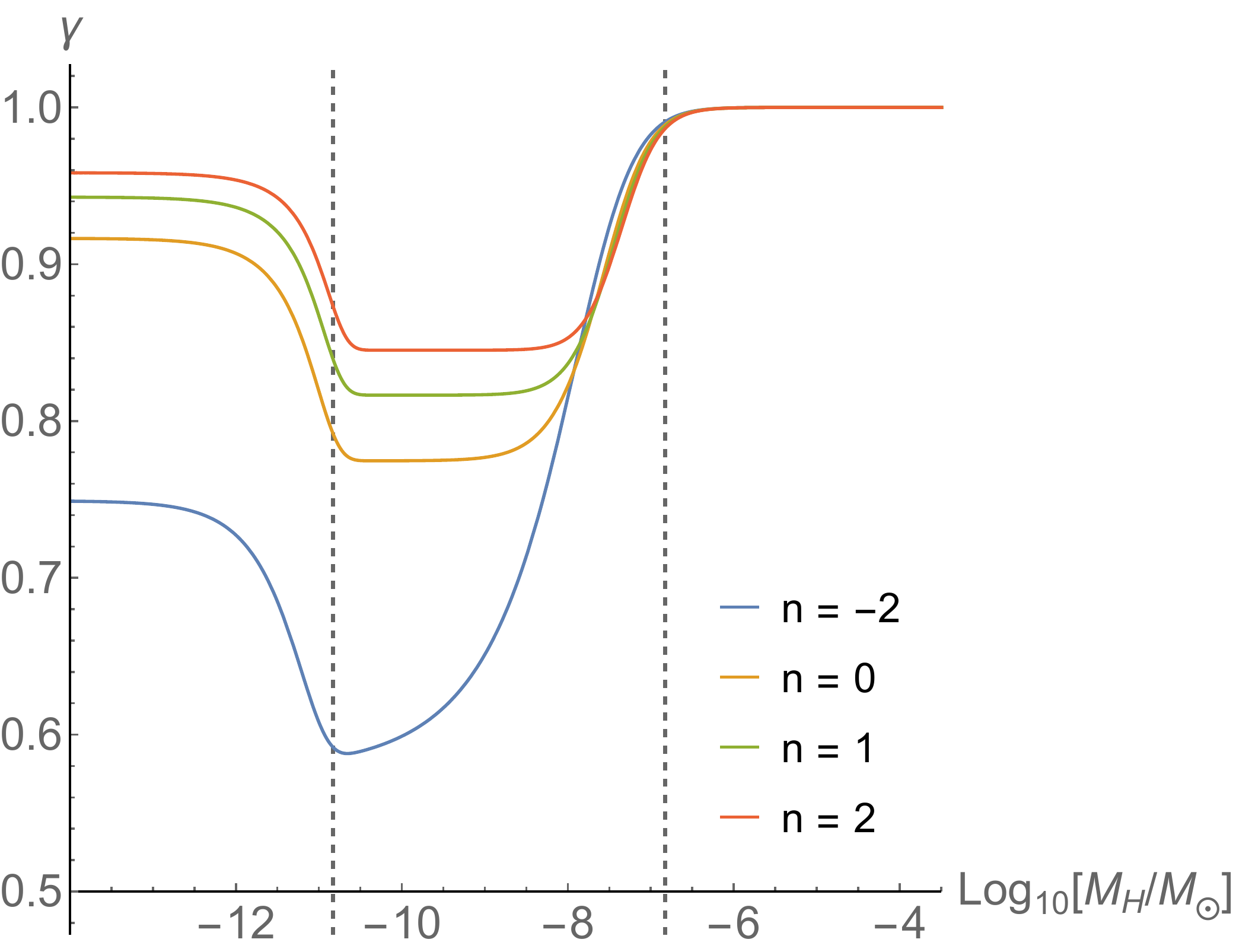}
	\end{center}
	\caption{
		\label{fig_tra}The variance $\sigma_\Delta (M_H)$ (left panel) and the $\gamma$ factor (right panel) of the trapezoidal template with the pivot scale $k_{\rm}$ corresponding to a horizon mass $ M_{H0} = 1.5\times 10^{-7} M_\odot$ and $k_{\rm max} = 100 k_{\rm min}$ (dashed lines).  
	}
\end{figure}

The important factor that featured the averaged spatial configuration around the peaks according to the input inflationary spectrum is
\begin{align}
\gamma = \frac{\sigma_1^2}{\sigma_\Delta \sigma_2}.
\end{align} 
This $\gamma$ factor enters the probability distribution of peaks when putting constraints on the spatial configuration.  We show examples for the $\gamma$ factor in terms of the horizon mass $M_H$ for the broken power-law template (in Figure~\ref{fig_BPL}) and the trapezoidal template (in Figure~\ref{fig_tra}). 
Note that the horizon mass $M_H$ can be translated into the smoothing scale through the useful relation
\begin{align}\label{horizon_mass to smoothing}
\frac{M_H}{M_{Heq}} = \left(k_{eq}R\right)^2 \left(\frac{g_{\ast eq}}{g_\ast}\right)^{1/3},
\end{align}
where $g_\ast$ is the number of relativistic degrees of freedom at the temperature of $M_H$, assuming to be the same as the entropy degrees of freedom. 
$M_{Heq} \approx 2.8 \times 10^{17} M_{\odot} \sim 10^{51}$g is the horizon mass at matter-radiation equality, where at this epoch $g_{\ast eq} \approx 3$ and $k_{eq} = 0.07\, \Omega_{m}h^2 \mathrm{Mpc}^{-1} \approx 0.01 \mathrm{Mpc}^{-1}$.

\section{Peak statistics with spatial constraints}
The selection of peaks as local maxima of the superhorizon density fluctuations 
imposes constraints on the spatial configuration at the local site of each peak. 
In order to take ensemble average based on these conditional peaks, one has to integrate over all relevant random variables constituted by the density perturbation and its first and second spatial derivatives. For 3-dimensional real space, the joint distribution in total involves with 10 random variables as shown in \cite{Bardeen:1985tr}. Here we focus on Gaussian random fields so that the probability distribution is completely fixed by the correlation among all variables. 


\subsection{general peak theory}
We first summarize the peak statistics of Bardeen, Bond, Kaiser and Szalay (BBKS) \cite{Bardeen:1985tr}, and for the application to PBH abundance we define the peak value of the density contrast as $\nu \equiv \Delta/\sigma_\Delta$, following the notation in \cite{Young:2014ana}. 
The number density of the maxima with height between $\nu$ and $\nu + d\nu$ is \cite{Bardeen:1985tr}:
\begin{align}\label{n_BBKS point process}
n_{\rm BBKS}(\mathbf{r}, \nu)d\nu = \sum_{p}^{} \delta^{(3)}(\mathbf{r}- \mathbf{r}_p),
\end{align}
where $\mathbf{r}_p$ are the positions of the local maxima.
{\footnote{The number density $n_{\rm BBKS}$ is independent of the position $\mathbf{r}_p$ due to the homogeneity of the density field.}}
 For each maximum, one can expand the point process as
\begin{align}
\delta^{(3)}(\mathbf{r}- \mathbf{r}_p) = \mathrm{det}\left\vert \mathbf{z}(\mathbf{r}_p) \right\vert \delta^{(3)}[\mathbf{u}(\mathbf{r})],
\end{align}
where $z_{ij}\equiv \partial_i\partial_j\Delta(\mathbf{r})$ are components of the second derivative tensor $\mathbf{z}$ and $u_i = \sum_j z_{ij}(r - r_p)_j$ are components of the gradient vector $\mathbf{u}$. Note that $z_{ij}$ is symmetric and has only six independent components.
The number density based on the point process \eqref{n_BBKS point process} has a dimension same as $\delta^{(3)}(\mathbf{r}- \mathbf{r}_p)$.

The conditions for $\Delta(\mathbf{r}_p)$ to be a local maximum ask $u_i(\mathbf{r}_p) =0$ and $\mathbf{z}(\mathbf{r}_p)$ to be negative definite. Assuming that $\Delta$ is Gaussian, the joint distributions of $u_i$ and $z_{ij}$ are also Gaussian. Therefore, to compute the number density of peaks we are in fact deal with a 10-dimensional random-field system with the probability distribution for 10 variables given by
\begin{align}\label{P_BBKS}
P_{\rm BBKS}(\Delta, u_i, z_{ij}) = \frac{\exp(-B)}{\sqrt{(2\pi)^{10} \mathrm{det}\vert \mathcal{M}\vert}},
\end{align}
where $B \equiv \sum \frac{1}{2} \Delta y_i (\mathcal{M}^{-1})_{ij}\Delta y_j$ with $\mathcal{M}_{ij}$ the covariance matrix, and $\Delta y_i = y_i - \langle y_i \rangle$ (for $i = 1,\cdots, 10$). Here we choose $y_1 = \Delta$, $y_i = u_i$ for $i = 2,3,4$, and $y_i = (z_{jk})_i$ for $i = 5,\cdots,10$ and $jk = 11,22,33,23,13,12$ for components of $\mathbf{z}$. We have restricted ourselves  to the zero mean setup $\langle \Delta\rangle = 0$.

The homogeneity and the isotropy of the underlying random field $\Delta$ allow us to integrate out all the spatial-dependent variables $y_i$, for $i = 2,\cdots, 10$, leading to an one-dimensional effective result $n_{\rm BBKS} = n_{\rm BBKS}(\nu)$. However, at best eight out of the 10 dimensions in the matrix $\mathcal{M}$ can be diagonalized by aligning $y_5$, $y_6$, $y_7$ to the principal axes of $\mathbf{z}$. The E-to-M condition ask eigenvalues of the principal axes to have non-positive values. 
There are off-diagonal terms arising from the non-vanished correlation between $\Delta$ and $z_{ij}$, whose effects on the PBH abundance is studied in the next section.
The detail of the dimensional reduction process $P(\Delta, u_i, z_{ij}) \rightarrow P(\Delta)$ is given in the Appendix A of \cite{Bardeen:1985tr}.

It is convenient to introduce the differential formula $n_{\rm BBKS}(\nu) = \int_{\nu}^{\infty} N_{\rm BBKS}(\nu^\prime)d\nu^\prime$, especially for computing the extended mass function via the number density (see Section~\ref{sec:mass_function}). 
We quote the one-dimensional expression from \cite{Bardeen:1985tr} as
\begin{align}\label{N_BBKS}
N_{\rm BBKS}(\nu) d\nu = \frac{1}{(2\pi)^2R_\ast^3}G(\gamma, \nu)e^{-\nu^2/2}d\nu,
\end{align}
where $\gamma = \sigma_1^2/(\sigma_2\sigma_\Delta)$ and $R_\ast = \sqrt{3}\sigma_1/\sigma_2$ are factors depending on the input inflationary spectrum. 
Note that $N_{\rm BBKS}$ has the dimension of $R^{-3}$ in this definition. 
The E-to-M constraint from the derivatives of the maxima implicitly encoded in the function
\begin{align}\label{G_general}
G(\gamma, \nu) = \int_{0}^{\infty}dx\frac{f(x)}{\sqrt{2\pi (1-\gamma^2)}}\exp\left[\frac{-(x-\gamma\nu)^2}{2(1-\gamma^2)}\right],
\end{align}
where the $\nu$-dependence in this function is the consequence of the non-zero correlation between $\Delta$ and the variable $x$.
{\footnote{To apply the result of \eqref{N_BBKS}, the $z_{ij}$ matrix has been diagonalized and the definition of the variable in $f(x)$ is $x\equiv -(z_{11}+z_{22}+z_{33})/\sigma_2$.}}
 The function $f(x)$ is
\begin{align}
f(x) &= \frac{x^3-3x}{2}\left[\mathrm{erf}\left(x\sqrt{\frac{5}{2}}\right)+\mathrm{erf}\left(\frac{x}{2}\sqrt{\frac{5}{2}}\right)\right] \\\nonumber
+& \sqrt{\frac{2}{5\pi}}\left[\left(\frac{31x^2}{4}+\frac{8}{5}\right)e^{-\frac{5x^2}{8}}+\left(\frac{x^2}{2}-\frac{8}{5}\right)e^{-\frac{5x^2}{2}}\right].
\end{align} 
If $G(\gamma, \nu)$ were just a constant, then the integration of $N_{\rm BBKS}$ with respect to $\nu$ shares a similar form as the Press-Schechter method (up to the spectrum-dependent factor $R_\ast$) and one expects $n_{\rm BBKS}$ has a peak value at the lower limit of integration $\nu = \nu_c$.
In general, however, $G(\gamma, \nu)$ can shift the peak value of $n_{\rm BBKS}$ away from the Press-Schechter prediction. We provide the numerical check of this discussion in the next section.

In summary, the homogeneity and the isotropy of the random fields allow us to perform the integration over $y_5$, $y_6$, $y_7$ as
\begin{align}
\nonumber\label{n_BBKS}
n_{\rm BBKS} 
&= \left\langle \text{det}\vert \textbf{z}\vert\delta^{(3)}(\textbf{u})  \Theta(-y_5)\Theta(-y_6)\Theta(-y_7) \Theta(\nu-\nu_c)\right\rangle, \\
&= \frac{1}{(2\pi)^2R_\ast^3}\int_{\nu_c}^{\infty} G(\gamma, \nu)e^{-\nu^2/2}d\nu,
\end{align}
where $y_5 = z_{11}$, $y_6 = z_{22}$, $y_7 = z_{33}$ and $z_{ii}$ have been fixed with the eigenvalues of $\mathbf{z}$ so that $z_{ij} = 0$ for $i \neq j$. The density fraction of the Universe that collapsed to form PBHs at the scale $R$ is estimated by $\beta_{\rm PBH} = V(R) n_{\rm peaks}$ \cite{Young:2014ana}, where $V(R)= (\sqrt{2\pi}R)^3$ is the volume of the Gaussian window function that satisfies the normalization condition $V(R)^{-1}\int W(x, R)d^3x =1$ with $W(x, R) = \exp[-x^2/(2R^2)]$ in the real space. 
It is remarkable that when any two of the eigenvalues are degenerate (such as in the case with exact spherical symmetry), the number of independent variables are reduced and one shall construct a lower-dimensional joint distribution of \eqref{P_BBKS}.

\subsection{point peak theory}
In this section we introduce a special peak statistics as a bridge to connect the general peak theory \cite{Bardeen:1985tr} and the so-called Press-Schechter (PS) method for the PBH abundance conventionally described by the Carr's formula \cite{Carr:1975qj}. The special peak theory is basically a two-step reduction of the BBKS method. The first step is to impose exact spherical symmetry to the system which reduces the number of independent random variables. The second step is to treat each selected peak as a dimensionless point-like object which reproduce the correct dimension of the PS method for the PBH abundance. 
The dimensionless reduction of the density peaks is nevertheless an unphysical approach.
We note that the first step is an intermediary process convenient for the discussion and, however, the system is expected to be spherically symmetric after the dimensionless point process. 

	Before invoking the spherical symmetry, we recall the variables relevant to the second spatial derivatives in the BBKS method \cite{Bardeen:1985tr} as
	\begin{align}
	z_1 &= -\partial^2 \Delta(\mathbf{r})/\sigma_2 = -(z_{11} + z_{22} +z_{33})/\sigma_2, \\\nonumber
	z_2 &= -(z_{11} - z_{33})/(2\sigma_2), \qquad z_3 = - (z_{11} - 2z_{22} + z_{33})/(2\sigma_2).
	\end{align}
	This definition maximally diagonalizes the covariance matrix $\mathcal{M}$ of the BBKS formalism \eqref{P_BBKS} for an arbitrary choice of axes with the only non-vanished correlations given as $\langle \nu^2\rangle = \langle z_1^2 \rangle =1$, $\langle \nu z_1\rangle = \gamma$, and $\langle z_2^2\rangle = \langle z_3^2\rangle/3 =1/15 $.
	The $B$ factor in the probability distribution function \eqref{P_BBKS} is rewritten as
	\begin{align}
	2B = \nu^2 + \frac{(z_1 - \gamma \nu)^2}{1- \gamma^2} + 15 z_2^2 + 5z_3^2 
	+ \frac{3\mathbf{u}\cdot\mathbf{u}}{\sigma_1^2} + \sum_{i = 8}^{10}\frac{15 y_i^2}{\sigma_2^2}.
	\end{align}
	The 10 variables are now $y_1 = \nu$, $y_i = u_i$ for $i = 2, 3, 4$, $y_5 = z_1$, $y_6 = z_2$, $y_7 = z_3$ and $y_8 = z_{23}$, $y_9 = z_{13}$, $y_{10} = z_{12}$.  
	The volume element involved with the six variables of the second spatial derivatives ($y_i$ for $i = 5,\cdots, 10$) is nothing but the volume element of the symmetric matrix $\mathbf{z}$, which can be expressed by \cite{Bardeen:1985tr}:
	\begin{align}
	d V_{\mathbf{z}} &= \prod_{i = 5}^{10} d y_i \\\nonumber
	&= \vert (\lambda_1 - \lambda_2)(\lambda_2- \lambda_3)(\lambda_1- \lambda_3)\vert d\lambda_1 d\lambda_2 d\lambda_3 \frac{d\Omega_3}{6}.
	\end{align}
	Here in the second equality the axes are chozen such that $z_i = -\lambda_i$ for $i = 1, 2, 3$ and $y_i = 0$ for $i = 8, 9, 10$ where $\lambda_i$ are  eigenvalues of $\mathbf{z}$. $d\Omega_3$ is the volume element of the three-dimensional rotation group $SO(3)$ and can be integrated out readily since $B$ is independent of the Euler angles.  
	In the case with exact spherical symmetry, we have identical eigenvalues $\lambda_1 = \lambda_2 = \lambda_3$ so that $d V_{\mathbf{z}} \rightarrow 0$.
	The volume of $\mathbf{z}$ collapses to a point as there is only one non-vanished random variable $z_1$ for the second spatial derivatives. 
	Similarly, there is only one random variable for the first derivatives with respect to spherical symmetry.

Let us now derive the probability distribution for the density peaks with perfect spherical symmetry. 
A local density peak smoothed by a window function $W(kR)$ in the high frequency limit is given by
\begin{align}
\Delta_R(\vert\textbf{r} - \textbf{r}_p\vert, R) \equiv \Delta_R(r, R) = \frac{1}{(2\pi)^3}\frac{4\pi}{r}\int_{0}^{\infty}dk k \sin (kr) \Delta_k(t) W(kR),
\end{align} 
where we have integrated out the angular dependence in this expansion.
Note that we do not label $r$ with respect to a specific peak position $\mathbf{r}_p$ as the ensemble average over selected peaks will finally become independent of the peak positions, given the homogeneity of the density field.

Our assumption is that the radial derivatives, $\partial_r\Delta_R$, $\partial_r\partial_r \Delta_R$, $\cdots$, are differentiable around the origin of the local peaks where $r \rightarrow 0$. In this limit, the correlation functions for the random fields of our interest are
\begin{align}
\frac{1}{r^2}\langle \partial_r\Delta_R\partial_r\Delta_R\rangle &=\frac{1}{9}\sigma_2^2,  \quad  
\frac{1}{r}\langle\Delta_R\partial_r\Delta_R\rangle = -\frac{1}{3}\sigma_1^2, \quad
\frac{1}{r}\langle \partial_r\Delta_R\partial_r^2\Delta_R\rangle =\frac{1}{9}\sigma_2^2, \\
&\langle \partial_r^2\Delta_R\partial_r^2\Delta_R\rangle =\frac{1}{9}\sigma_2^2, \quad
\langle \Delta_R\partial_r^2\Delta_R\rangle = -\frac{1}{3}\sigma_1^2.
\end{align}
One can observe that all correlations involved with $\partial_r\Delta_R$ vanish explicitly at $r = 0$, where the extremum constraint $\partial_r\Delta_R = 0$ holds in an apparent way and the maximum condition $\partial_r^2 \Delta_R < 0$ only introduces one more statistical variable in addition to $\Delta_R$.


We now impose the dimensionless point process by fixing $r = 0$. This process may be intuitively illustrated by a dimensional reduction of \eqref{n_BBKS point process} with $\delta^{(3)}(\textbf{r} - \textbf{r}_p)\rightarrow \delta^{(0)}(r)$. In other words we are now collecting point-like peaks with zero dimension in space.  
Following the normalization for the two variables $\nu = \Delta_R/\sigma_\Delta$ $z_r = - \partial_r^2\Delta_R/\sigma_2$, the covariance matrix of the two-dimensional statistical system reads
\begin{align}\label{covariant matrix_spk}
\mathcal{M}_{ij} = 
\begin{bmatrix}
1 &  \gamma/3 \\
\gamma/3 & 1/9
\end{bmatrix},
\end{align}
where both variables have zero means $\langle \nu\rangle = \langle z_r \rangle = 0$.
The joint probability distribution is then given by
\begin{align}
P_{\rm spk}(\nu, z_r)d\nu dz_r = \frac{e^{-B}}{\sqrt{(2\pi)^2\textrm{det}(\mathcal{M})}}d\nu dz_r. 
\end{align}
Taking $y_1 = \nu$ and $y_2 = z_r$, the $B$ factor can be computed by 
\begin{align}
2B = \sum y_i\left(\mathcal{M}^{-1}\right)_{ij}y_j = \nu^2 +\frac{(\gamma\nu - 3z_r)^2}{1-\gamma^2},
\end{align}
and we have imposed the result of \eqref{covariant matrix_spk} in the second equality. The probability function for the special peak theory is therefore
\begin{align}
P_{\rm spk} (\nu, z_r) = \frac{3}{2\pi}\frac{e^{-B}}{\sqrt{1-\gamma^2}}.
\end{align}

The number density of point-like peaks that satisfies the E-to-M conditions for PBH formation above the threshold $\nu_c$ is evaluated in the usual way as
\begin{align}\label{n_spk definition}
n_{\rm spk} &= 2\int_{-\infty}^{\infty} \int_{-\infty}^{\infty} P_{\rm spk}(\nu, z_r) \Theta(z_r) \Theta(\nu - \nu_c)  dz_r d\nu, \\
&= \frac{6}{2\pi} \int_{\nu_c}^{\infty} \int_{0}^{\infty} \frac{e^{-\nu^2/2}}{\sqrt{1-\gamma^2}}
\exp\left[-\frac{(\gamma\nu -3z_r)^2}{2(1-\gamma^2)}\right] dz_r d\nu, \\
\label{n_spk inter1}
&= \frac{2}{2\pi} \sqrt{\frac{\pi}{2}} \int_{\nu_c}^{\infty} e^{-\nu^2/2} 
\left(1+ \text{erf}\left[\frac{\gamma \nu}{\sqrt{2(1-\gamma^2)}}\right]\right) d\nu.
\end{align}
The factor $2$ in \eqref{n_spk definition} is introduced for an alignment with the PS method in the high peak limit.
By using the change of variable $x = \gamma\nu/\sqrt{2-2\gamma^2}$ so that $\nu^2 = x^2(2-2\gamma^2)/\gamma^2$, we can arrive at the analytical expression for $n_{\rm spk}$ as
\begin{align}\label{n_spk full result}
n_{\rm spk} = \frac{1}{2}\text{erfc}(\nu_c/\sqrt{2}) 
-2 \left[T_\infty\left(\sqrt{\frac{\gamma^2}{1-\gamma^2}}\right)- T_{\nu_c}\left(\sqrt{\frac{\gamma^2}{1-\gamma^2}}\right)\right],
\end{align}
where $T_{n}(x)$ is the Owen's T function with the definition
\begin{align}
T_{n}(x) = \frac{1}{2\sqrt{2\pi}}\int_{-n}^{\infty}e^{-t^2/2}\text{erf}\left(\frac{xt}{\sqrt{2}}\right) dt.
\end{align}
The dependence on the $\gamma$ factor in the $T_{n}(x)$ functions reflects the effect of E-to-M constraints on the number density of peaks. 
Given that $n_{\rm spk}$ is dimensionless in this definition,
the PBH density is simply $\beta_{\rm spk} = n_{\rm spk}$.

\subsection{The high peak expansion}
For $0< \gamma< 1$, the $G(\gamma, \nu)$ function as the integrand of \eqref{n_BBKS} in general relies on numerical computation. In the limit of $\gamma \nu \gg 1$, the function $\exp [-(x -\gamma\nu)^2/2(1-\gamma^2)]$ behaves as a delta function in $G(\gamma , \nu)$ so that $f(x)$ picks up the value around $x = \gamma\nu$. This leads to the asymptotic expansion in the large $\gamma\nu$ limit as \cite{Bardeen:1985tr}:
\begin{align}\label{G_highpeak}
G(\gamma, \nu) \rightarrow (\gamma\nu)^3 - 3(\gamma\nu).
\end{align}
The high peak expansion thus gives rise to the analytical expression of \eqref{n_BBKS}, in terms of the dimensionless parameter $\beta_{\rm BBKS} = V(R)n_{\rm BBKS}$, as
\begin{align}\label{n_BBKS_high}
\beta_{\rm BBKS} = \frac{1}{\sqrt{2\pi}} \left[\left(\frac{R\sigma_1}{\sqrt{3}\sigma_\Delta}\right)^3(2+\nu_c^2) - \frac{R^3\sigma_2^2}{\sqrt{3}\sigma_1\sigma_\Delta}\right]e^{-\nu_c^2/2}+ \cdots,
\end{align}
where this expansion breaks down if $(\gamma\nu)^3 < 3(\gamma\nu)$. 

\begin{figure}
	\begin{center}
		\includegraphics[width=70mm]{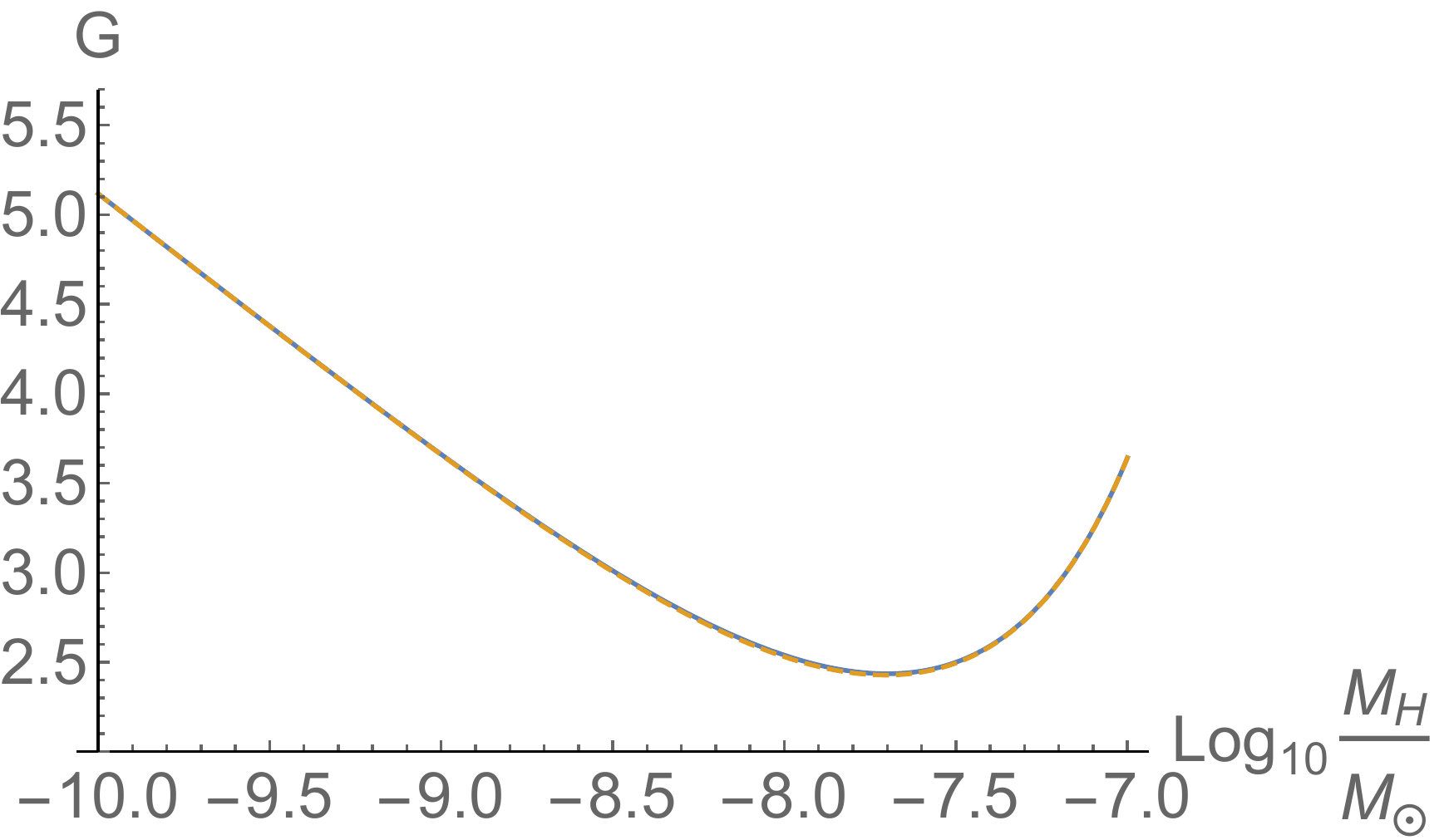}
		\hfill
		\includegraphics[width=70mm]{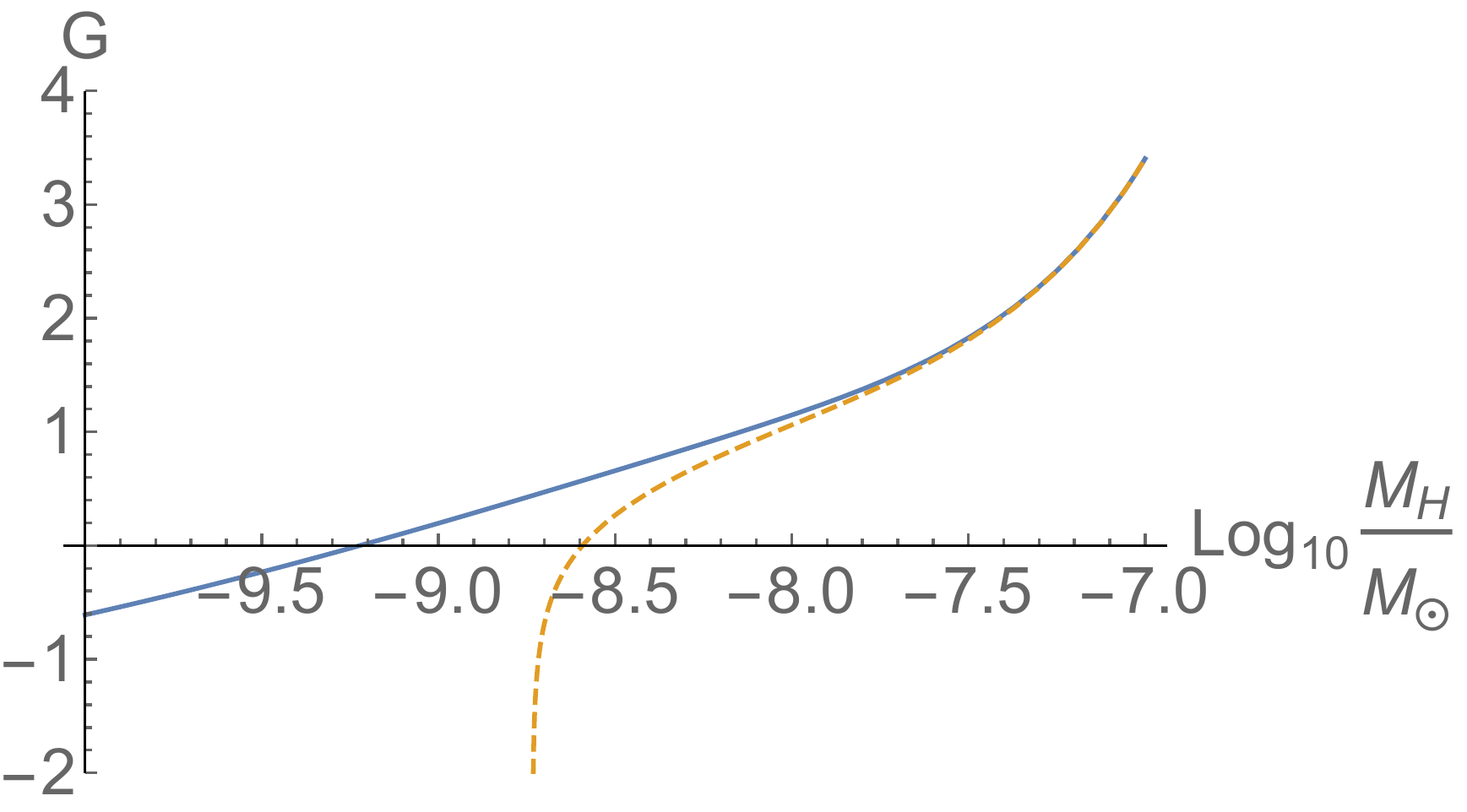}
	\end{center}
	\caption{
		\label{fig_G_BPL}The numerical results (solid lines) and the high peak approximation (dashed lines) of the function $G(\gamma, \nu)$ by using the broken power-law template with the pivot scale $k_0$ corresponding to a horizon mass $ M_{H0} = 1.5\times 10^{-7} M_\odot$. The spectral indices are chosen as $n =-1$ (left panel) and $n = 2$ (right panel). 
	}
\end{figure}

\begin{figure}
	\begin{center}
		\includegraphics[width=70mm]{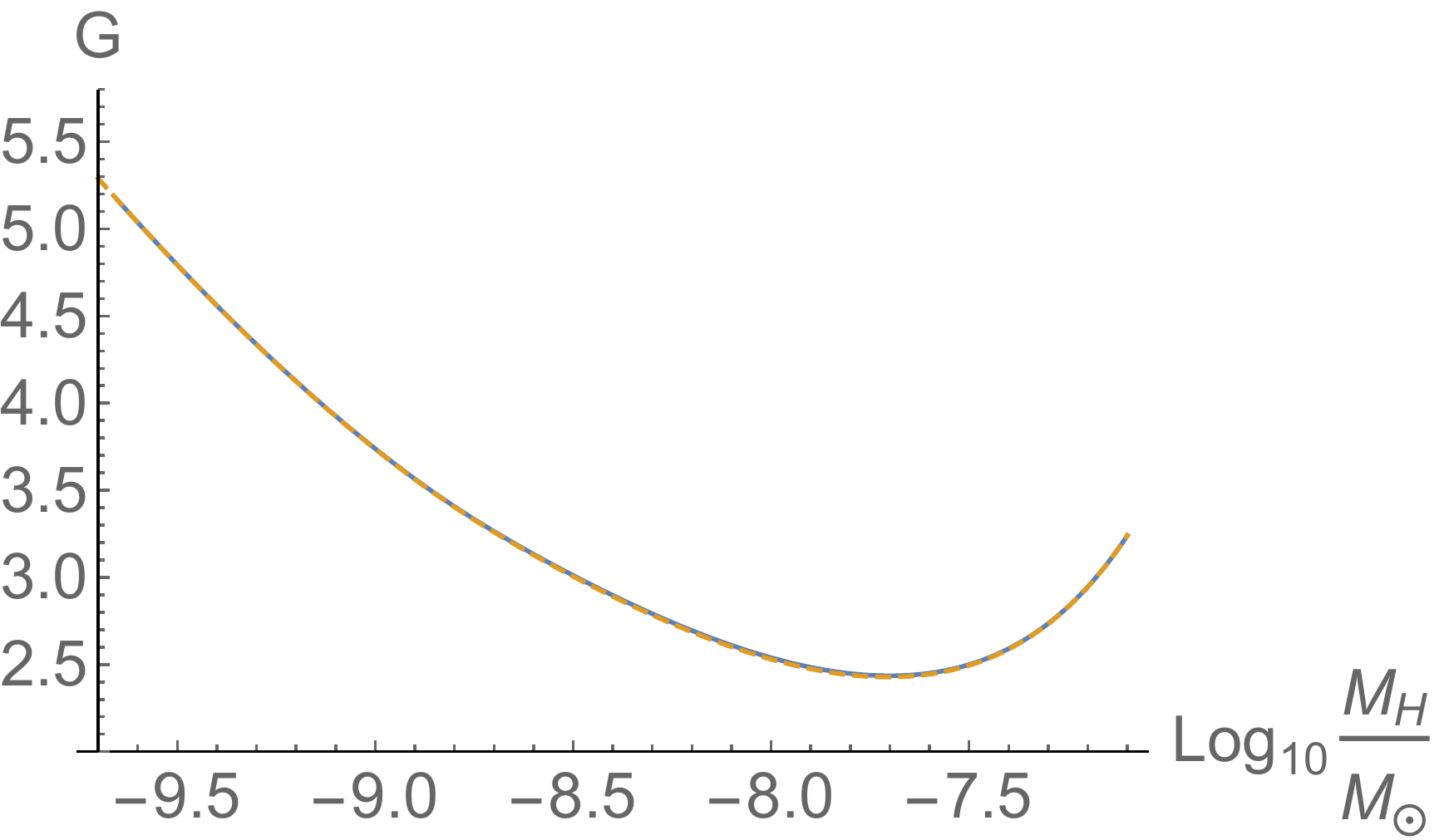}
		\hfill
		\includegraphics[width=70mm]{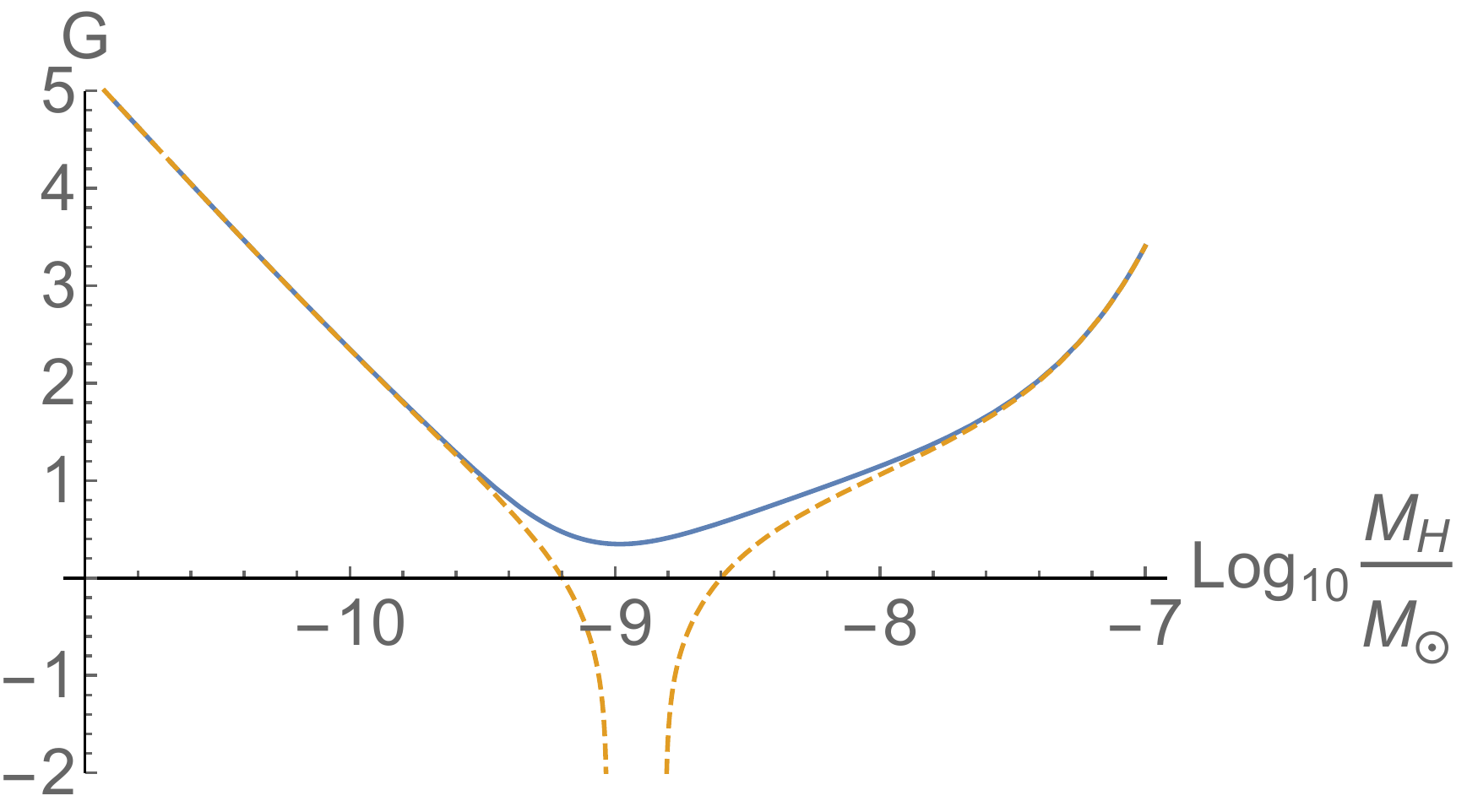}
	\end{center}
	\caption{
		\label{fig_G_tra}The numerical results (solid lines) and the high peak approximation (dashed lines) of the function $G(\gamma, \nu)$ by using the trapezoidal template with the pivot scale $k_{\rm min}$ corresponding to a horizon mass $ M_{H0} = 1.5\times 10^{-7} M_\odot$ and $k_{\rm max} = 10 k_{\rm min}$. The spectral indices are chosen as $n =-1$ (left panel) and $n = 2$ (right panel). 
	}
\end{figure}

We examine the validity of the high peak approximation \eqref{G_highpeak} with the broken power-law template in Figure~\ref{fig_G_BPL} and with the trapezoidal template in Figure~\ref{fig_G_tra}. Our results show that \eqref{G_highpeak} is a good approximation for both templates with $n \leq 1$. The high peak expansion of $G(\gamma, \nu)$ breaks down in the small mass limit for blue-tilted power-law spectrum with $n > 1$ and also in a range of the blue trapezoidal spectrum, depending on the ratio $k_{\rm max}/k_{\rm min}$. For trapezoidal spectra, the high peak approximation always holds in the small mass limit for $-2 \leq n \leq 2$.

On the other hand, one can apply the expansion of the error function in the large $x$ limit to \eqref{n_spk inter1} based on
\begin{align}
\text{erf}(ax) = 1- \frac{1}{a}\frac{1}{\sqrt{\pi}x} e^{-a^2x^2} +\cdots.
\end{align} 
This gives the high peak expansion ($\nu \gg 1$) of the number density \eqref{n_spk full result} with spherical symmetry as
\begin{align}
\beta_{\rm spk} = \beta_{\rm PS} 
- \frac{\sqrt{2}}{2\pi} \frac{\sqrt{1-\gamma^2}}{\gamma} \Gamma\left(0, \frac{\nu_c^2}{2(1-\gamma^2)}\right) + \cdots
\end{align}
where the first term $\beta_{\rm PS}$ is nothing but the PBH density according to the Press-Schechter method (i.e. the Carr's formula \cite{Carr:1975qj}):
\begin{align}\label{n_PS}
\beta_{\rm PS} = 2\int_{\nu_c}\frac{1}{\sqrt{2\pi}}e^{-\nu^2/2}d\nu = \text{erfc}(\frac{\nu_c}{\sqrt{2}}).
\end{align}
In the high peak limit $\nu_c \gg 1$, the PS number density reads
\begin{align}\label{n_PS_high}
\beta_{\rm PS} \rightarrow \sqrt{\frac{2}{\pi}}\frac{1}{\nu_c} e^{-\nu_c^2/2} + \cdots.
\end{align}
Comparing the leading terms of \eqref{n_BBKS_high} and \eqref{n_PS_high} in the limit of $\nu_c \gg 1$ one finds 
\begin{align}\label{high_peak_ratio}
\frac{\beta_{\rm BBKS}}{\beta_{\rm PS}} \sim \frac{1}{2}Q^{3/2} \nu_c^3,
\end{align}
where we denote $Q = R^2\sigma_1^2/(3\sigma_\Delta^2)$.
One can remove the factor $1/2$ in \eqref{high_peak_ratio} by supplying a factor $2$ to $\beta_{\rm BBKS}$ as what has been done for $\beta_{\rm PS}$ in \eqref{n_PS}. 
 The above relation was firstly examined in \cite{Young:2014ana} with blue-tilted power-law spectra which reports $\beta_{\rm BBKS}/\beta_{\rm PS} \sim \nu_c^3 \sim \mathcal{O}(10)$. However this result implies the break down of using the high peak expansion \eqref{n_BBKS_high} and $\beta_{\rm BBKS}$ generally can only be computed by numerical methods.

\subsection{Primordial black hole abundance}
We compare the PBH formation probability at each Hubble mass scale $M_H$ from different inflationary spectra. The prediction of the general peak statistics is $\beta_{\rm BBKS} = V(R) n_{\rm BBKS}$ with $n_{\rm BBKS}$ given by \eqref{n_BBKS}. The high peak approximation of $\beta_{\rm BBKS}$ is given by \eqref{n_BBKS_high}. We use \eqref{n_spk full result} for the special peak statistics $\beta_{\rm spk}$ (no high peak expansion). Our definition for the Press-Schechter formalism $\beta_{\rm PS}$ is given in \eqref{n_PS}.
We use $\Delta_c = 0.45$ as a typical value in the following demonstration. The precise value of $\Delta_c$ depends on the density profile of each peak, and the density profile should have correlation with the inflationary spectrum $\mathcal{P}_\zeta$. For the construction of a $\Delta_c$-$\mathcal{P}_\zeta$ correlated statistics, see \cite{Germani:2019zez,Kalaja:2019uju}.

\begin{figure}
	\begin{center}
		\includegraphics[width=70mm]{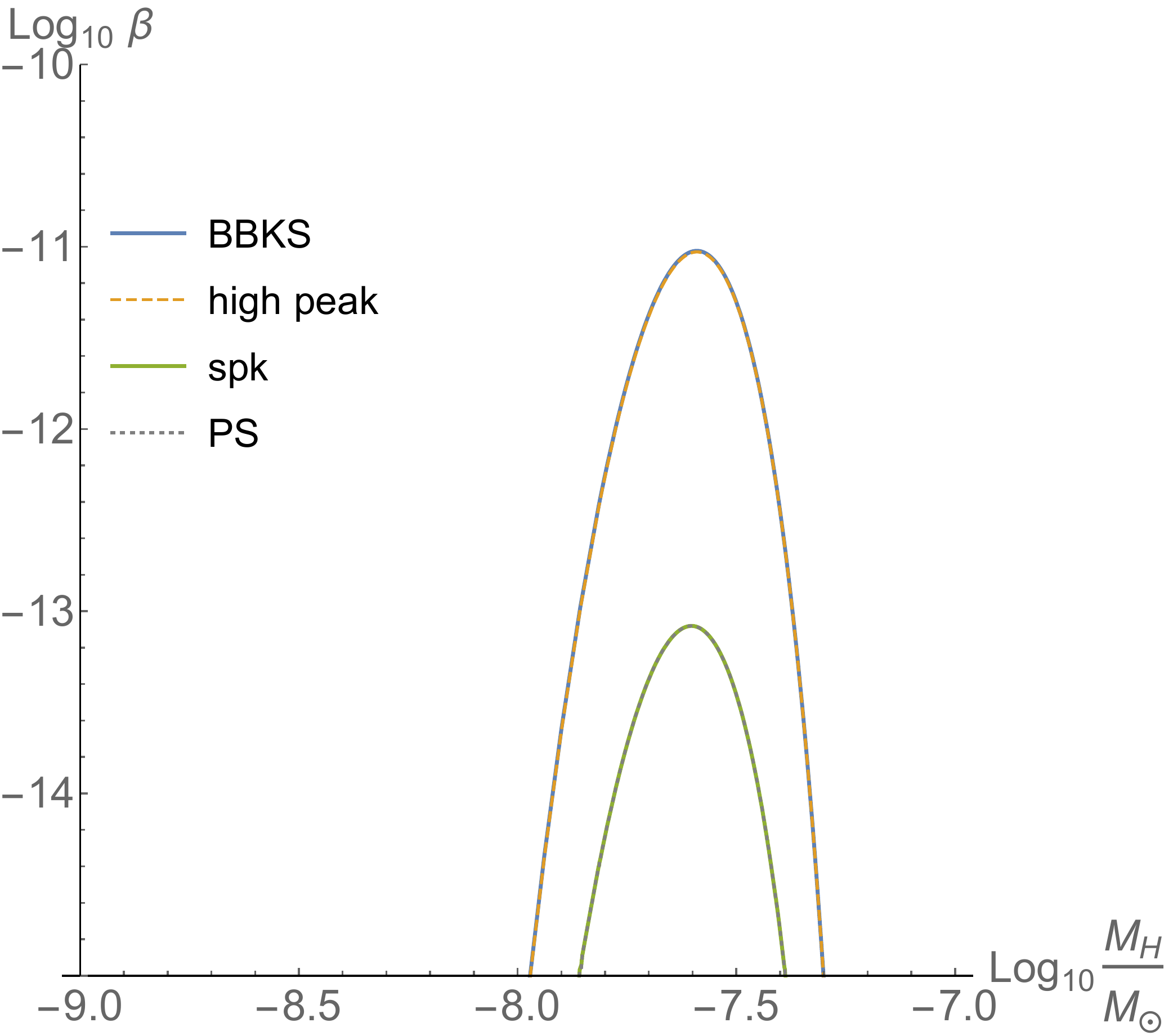}
		\hfill
		\includegraphics[width=70mm]{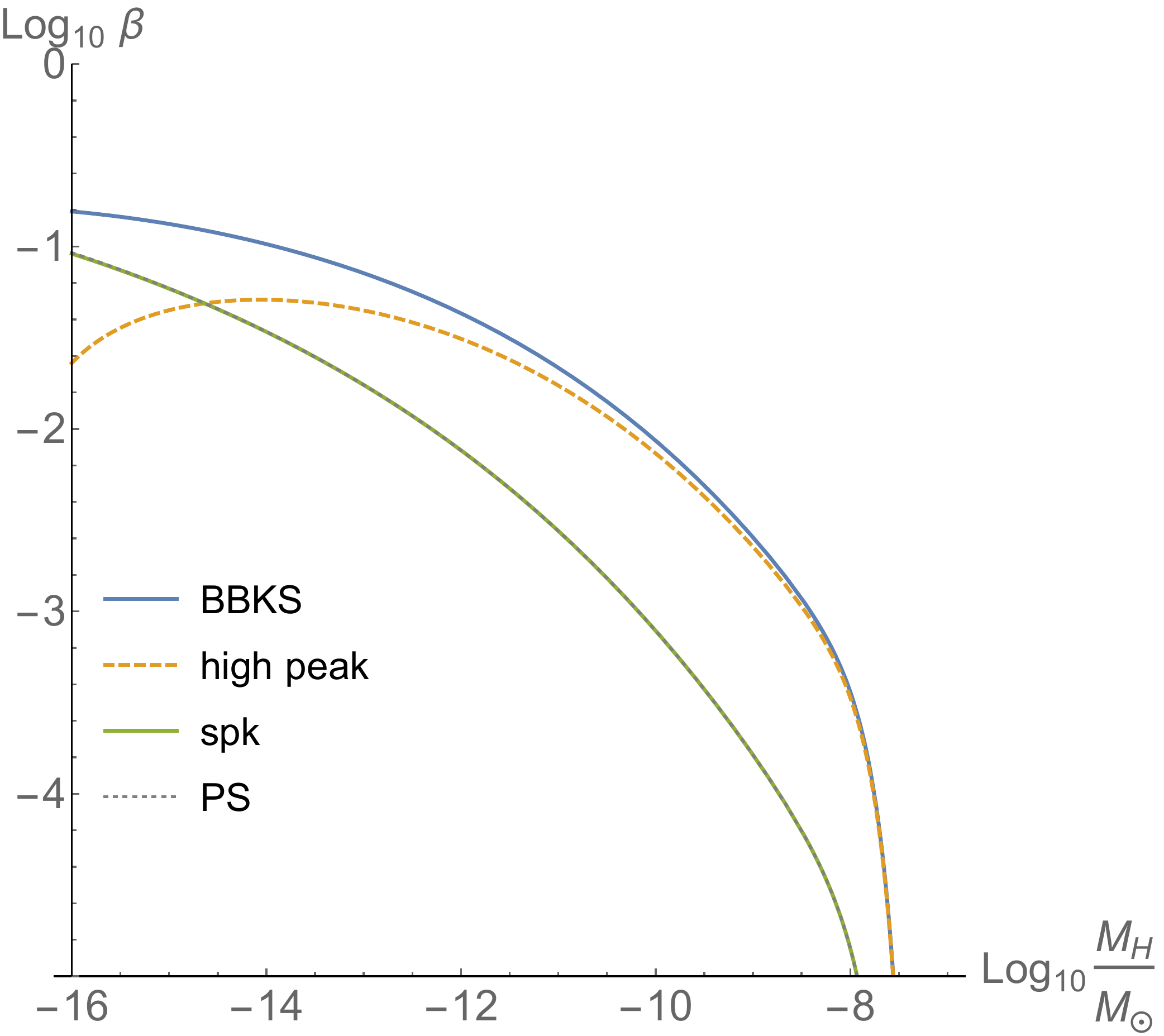}
	\end{center}
	\caption{
		\label{fig_beta_BPL}The PBH abundance $\beta (M_H)$ based on the broken power-law spectrum with the pivot scale $k_0$ corresponding to a horizon mass $ M_{H0} = 1.5\times 10^{-7} M_\odot$. The spectral indices are chosen as $n =-1$ (left panel) and $n = 1.2$ (right panel). 
	}
\end{figure}
\begin{figure}
	\begin{center}
		\includegraphics[width=70mm]{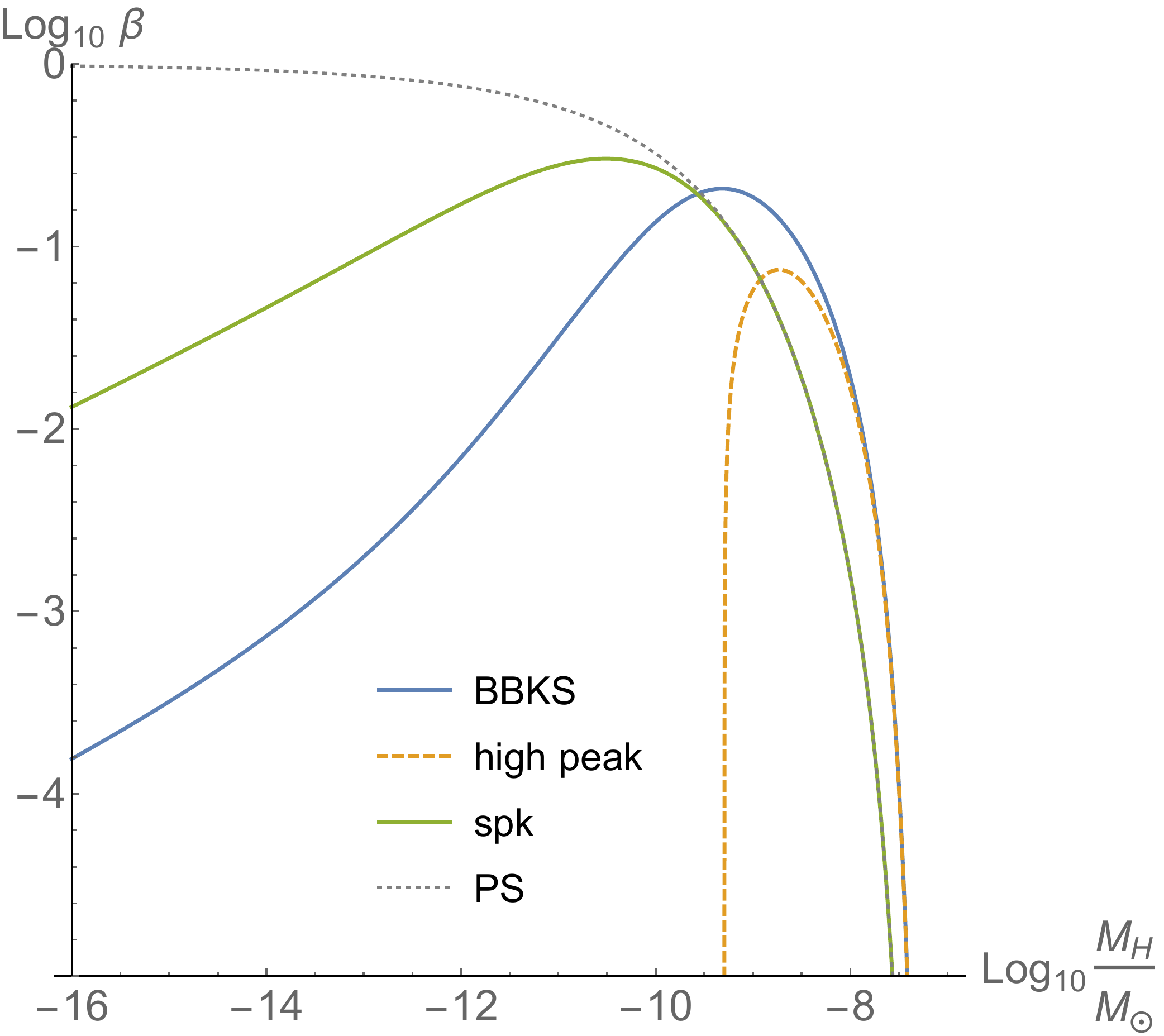}
		\hfill
		\includegraphics[width=70mm]{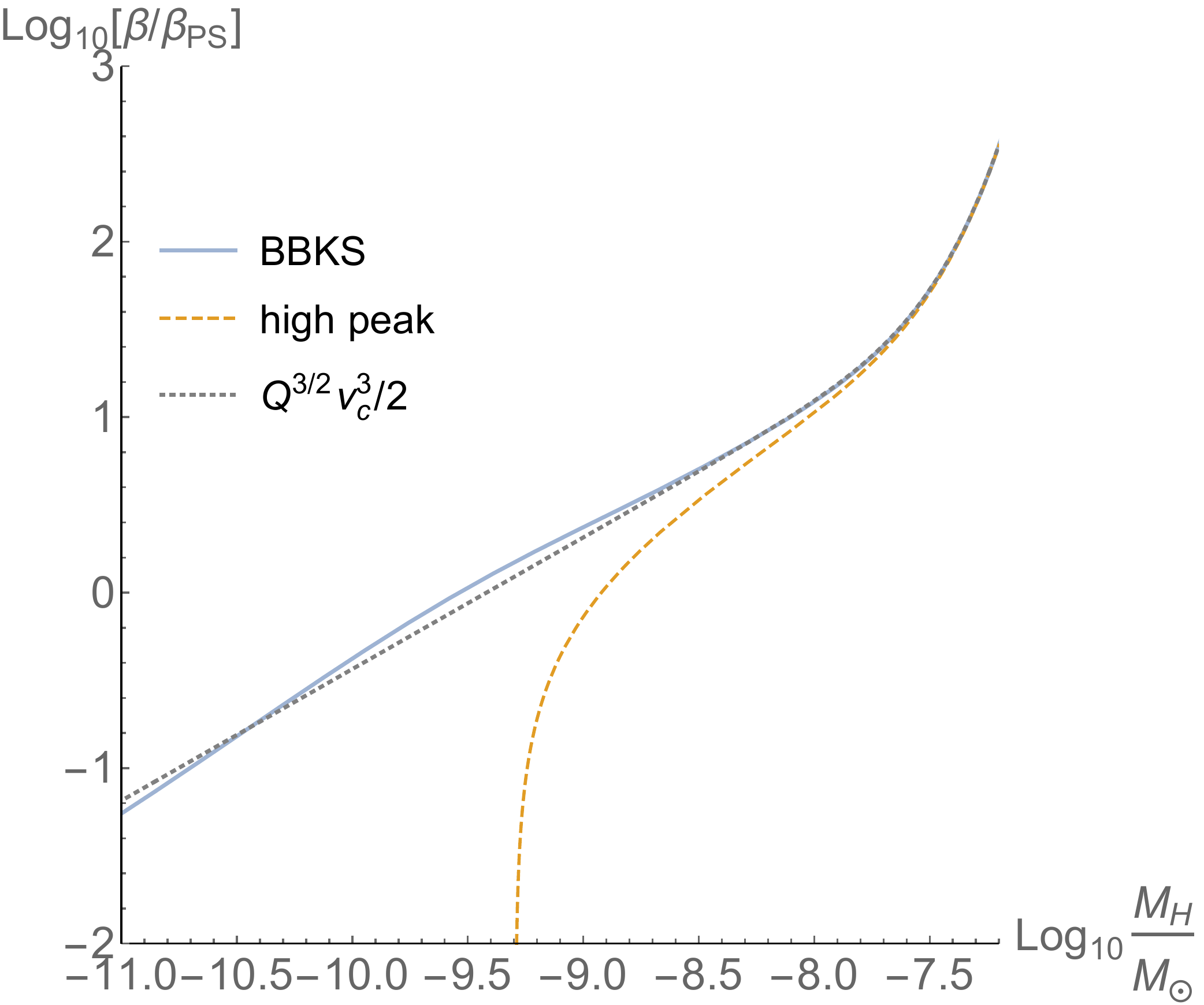}
	\end{center}
	\caption{
		\label{fig_beta_ratio_BPL}The PBH abundance $\beta (M_H)$ based on the broken power-law spectrum with the pivot scale $k_0$ corresponding to a horizon mass $ M_{H0} = 1.5\times 10^{-7} M_\odot$. The spectral indices are chosen as $n = 2$ (left panel) and the abundance ratio among statistical methods is given in the right panel. 
	}
\end{figure}

\smallskip
\noindent
\textbf{Broken power-law templates.}
For the spectral index $n \leq 1$, the high peak expansion of $\beta_{\rm BBKS}$ is a good approximation and $\beta_{\rm spk}$ coincides with $\beta_{\rm PS}$. Our numerical results show that the ratio $\beta_{\rm BBKS}/\beta_{\rm PS} = Q^{3/2}\nu_c^3/2 \sim 10^2$ for red-tilted templates.
For the spectral index $n > 1$ the high peak expansion breaks down in the small $M_H$ limit, as shown in Figure~\ref{fig_beta_BPL} and Figure~\ref{fig_beta_ratio_BPL}. For super blue-tilted templates ($n = 2$) $\beta_{\rm spk}$ start to deviate from $\beta_{\rm PS}$ in the small $M_H$ limit due to the important contribution from the $T$ function in \eqref{n_spk full result} when $\gamma$ approaches to $1$. For blue-tilted templates the ratio $\beta_{\rm BBKS}/\beta_{\rm PS}$ decreases with $M_H$ yet $Q^{3/2}\nu_c^3/2 $ still provides a good estimation for the difference in the resulting abundance.

The $Q$ factor for the broken power-law templates reads
\begin{align}
Q = \frac{1}{3}\frac{\Gamma\left(\frac{n+5}{2},k_0^2R^2\right)}{\Gamma\left(\frac{n+3}{2},k_0^2R^2\right)}.
\end{align}
One can numerically check that the $Q$ factor is a constant for $k_0 R\ll 1$ and $Q^{3/2} \leq 1$ for $-2 \leq n \leq 2$. For $k_0 R\gg 1$, $Q$ with different choices of $n$ converge to a same value and $Q^{3/2}$ can be much greater than $1$.

\begin{figure}
	\begin{center}
		\includegraphics[width=70mm]{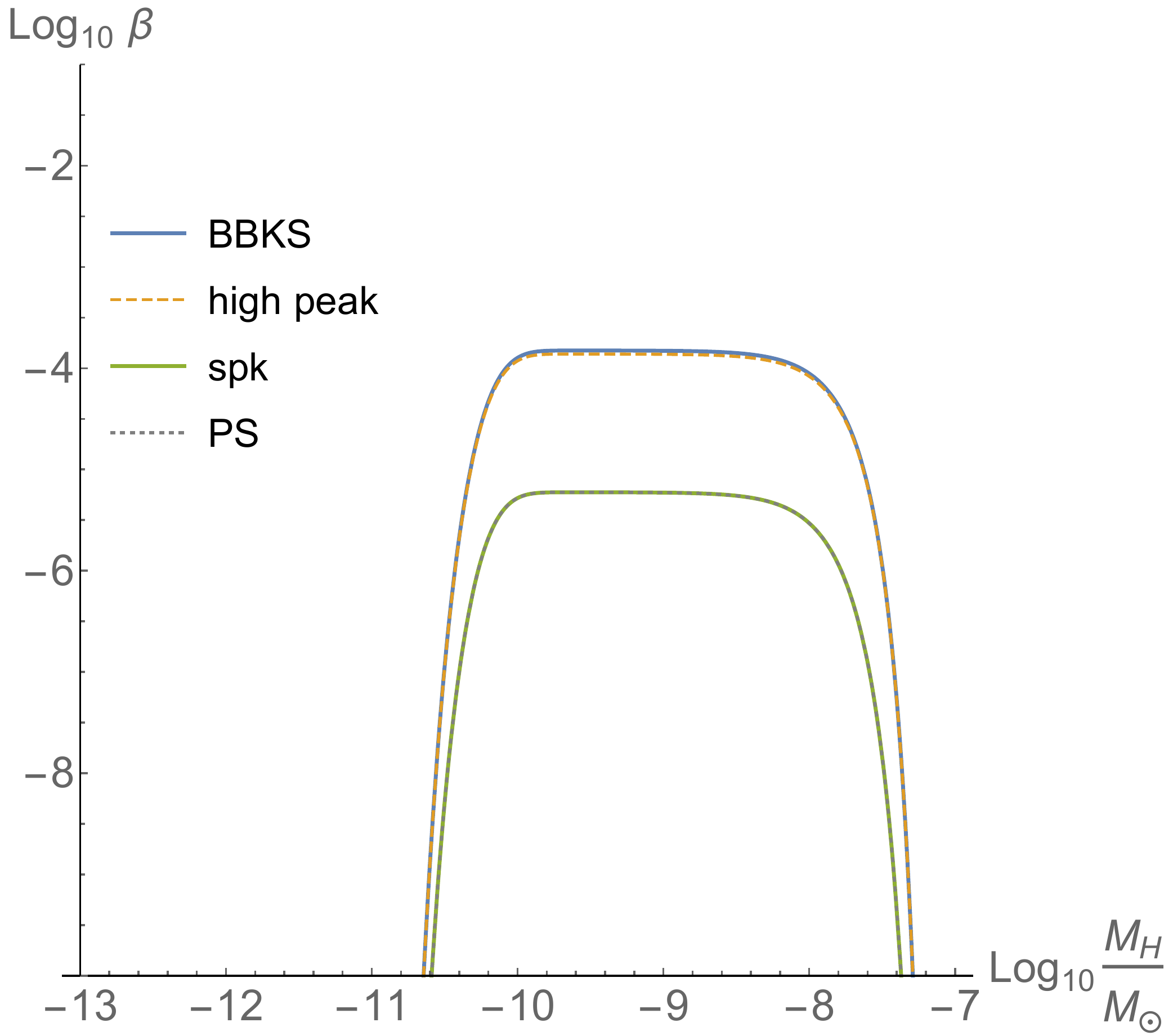}
		\hfill
		\includegraphics[width=70mm]{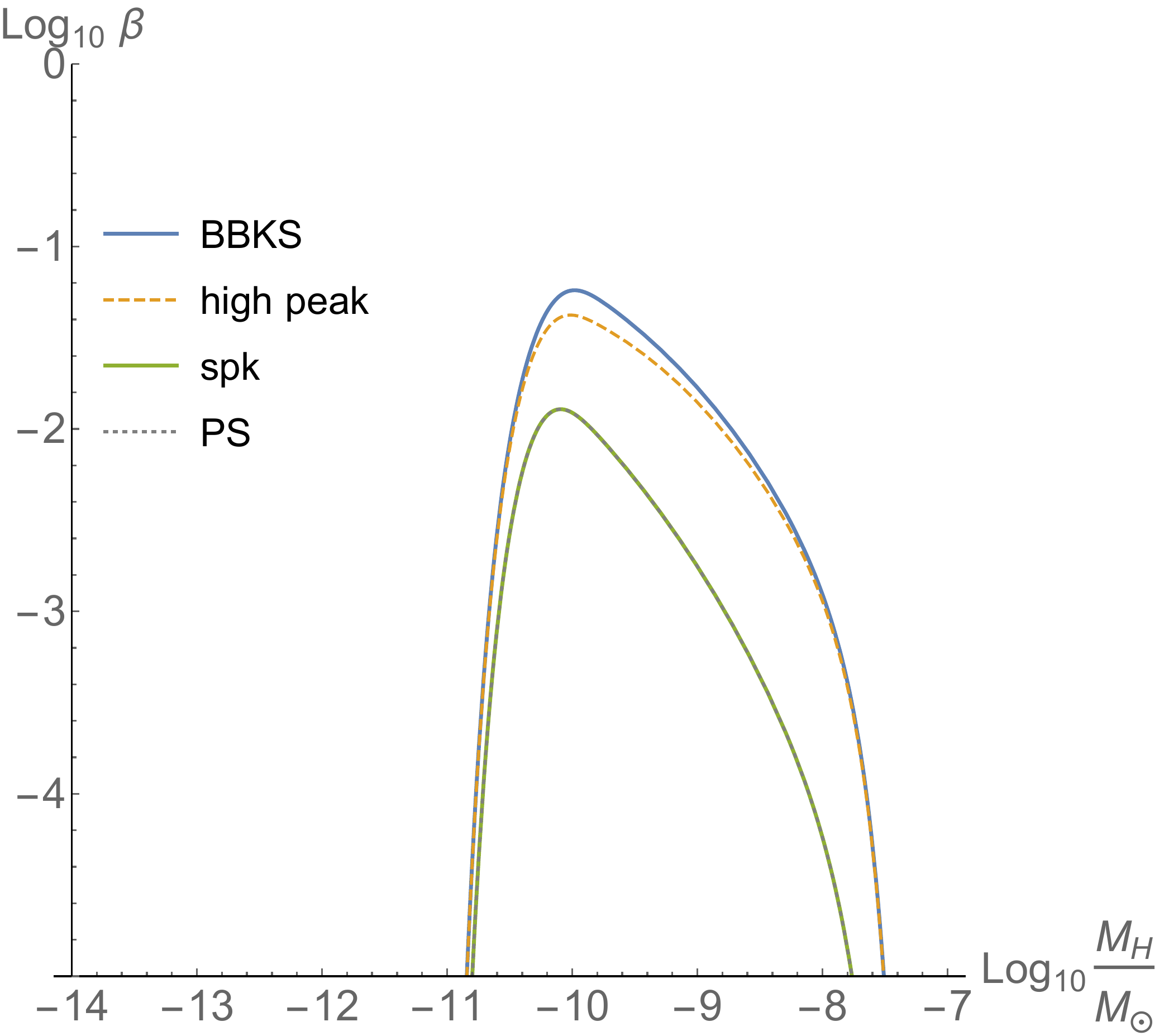}
	\end{center}
	\caption{
		\label{fig_beta_tra1}The PBH abundance $\beta (M_H)$ based on the broken power-law spectrum with the pivot scale $k_{\rm min}$ corresponding to a horizon mass $ M_{H0} = 1.5\times 10^{-7} M_\odot$ and $k_{\rm max} = 40 k_{\rm min}$. The spectral indices are chosen as $n = 1$ (left panel) and $n = 1.4$ (right panel). 
	}
\end{figure}
\begin{figure}
	\begin{center}
		\includegraphics[width=70mm]{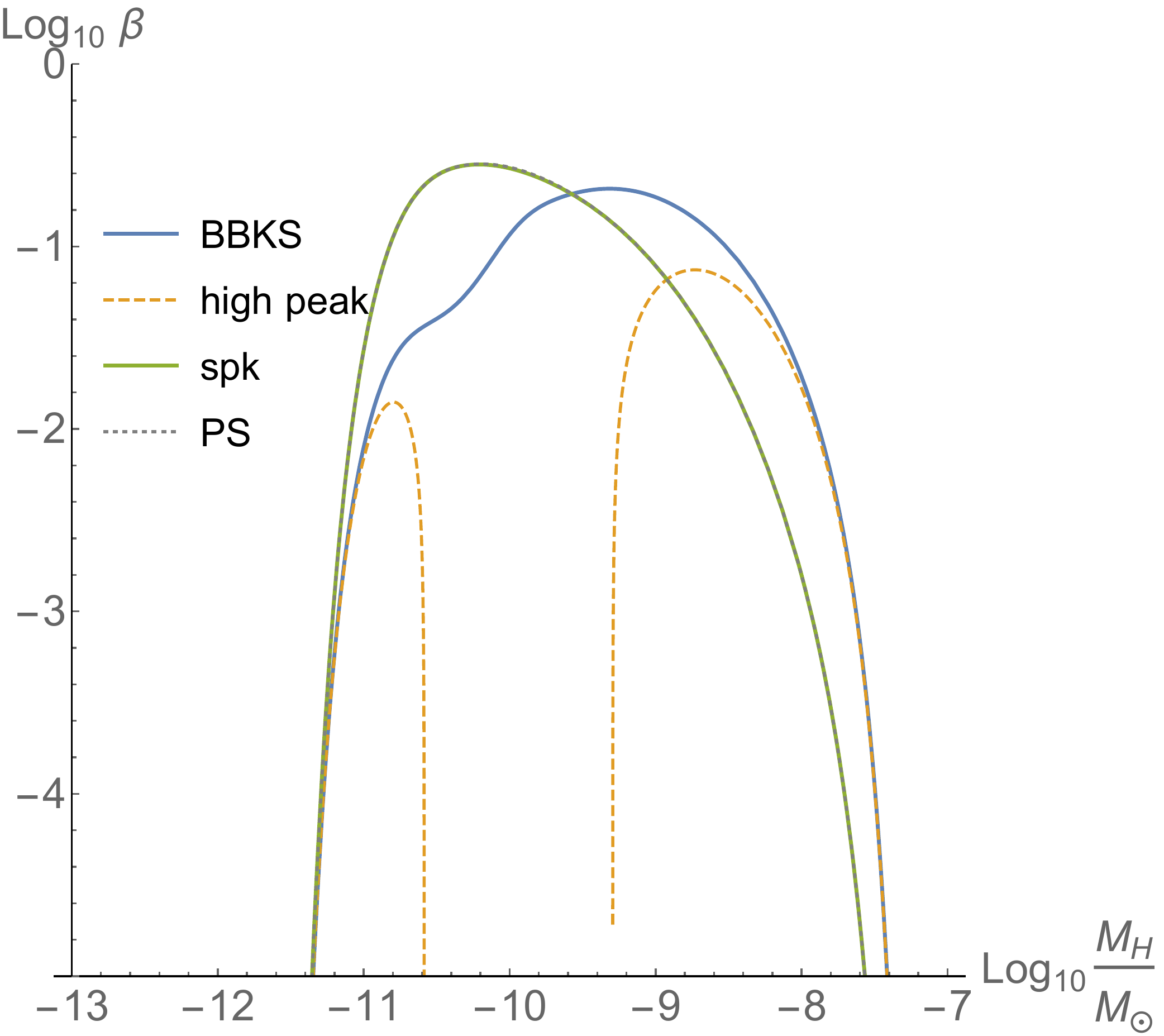}
		\hfill
		\includegraphics[width=70mm]{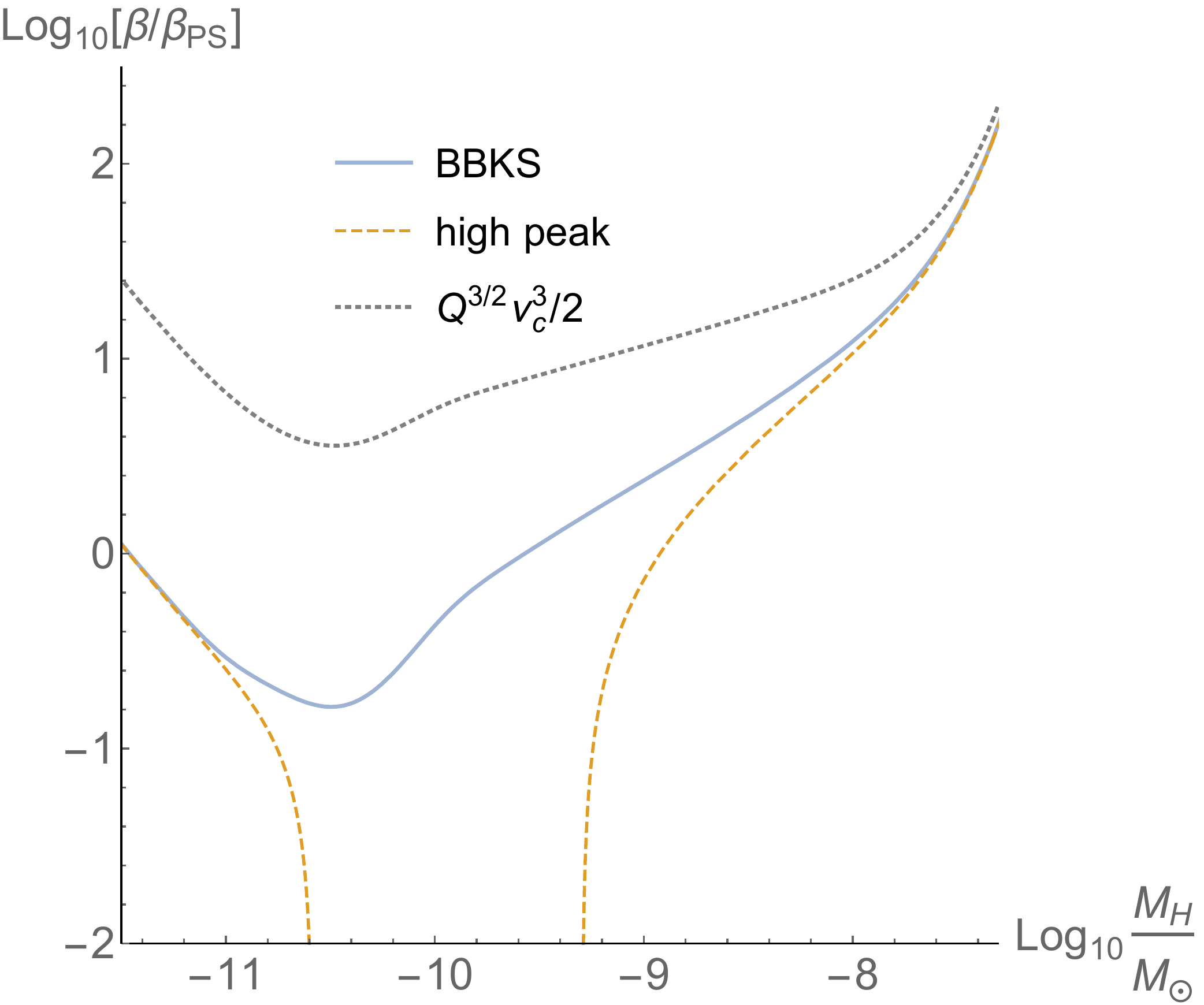}
	\end{center}
	\caption{
		\label{fig_beta_tra2}The PBH abundance $\beta (M_H)$ based on the broken power-law spectrum with the pivot scale $k_{\rm min}$ corresponding to a horizon mass $ M_{H0} = 1.5\times 10^{-7} M_\odot$ and $k_{\rm max} = 40 k_{\rm min}$. The spectral indices are chosen as $n = 2$ (left panel) and the abundance ratio among statistical methods is given in the right panel. 
	}
\end{figure}

\smallskip
\noindent
\textbf{Trapezoidal templates.}
For the spectral index $n = 1$ (top-hat), the high peak expansion of $\beta_{\rm BBKS}$ is a good approximation and $\beta_{\rm spk}$ coincides with $\beta_{\rm PS}$. The high peak expansion starts to deviate from $\beta_{\rm BBKS}$ in blue-tilted cases with $n > 1$. The numerical results of $\beta_{\rm PS}$ agree nicely with $\beta_{\rm spk}$ up to $n = 2$. For the blue-tilted cases $n > 1$ the high peak expansion breaks down between $k_{\rm min}$ and $k_{\rm max}$. $Q^{3/2}\nu_c^3/2 $ is a good estimation for the ratio $\beta_{\rm BBKS}/\beta_{\rm PS}$ when $n < 3/2$, and for super blue-tilted case with $n = 2$, $Q^{3/2}\nu_c^3/2 $ is no longer a good estimation for the ratio $\beta_{\rm BBKS}/\beta_{\rm PS}$. It is interesting to note that the peak value of $\beta_{\rm BBKS}$ is different from $\beta_{\rm PS}$ in the super blue-tilted case, as shown in Figure~\ref{fig_beta_tra2}.

\begin{figure}
	\begin{center}
		\includegraphics[width=70mm]{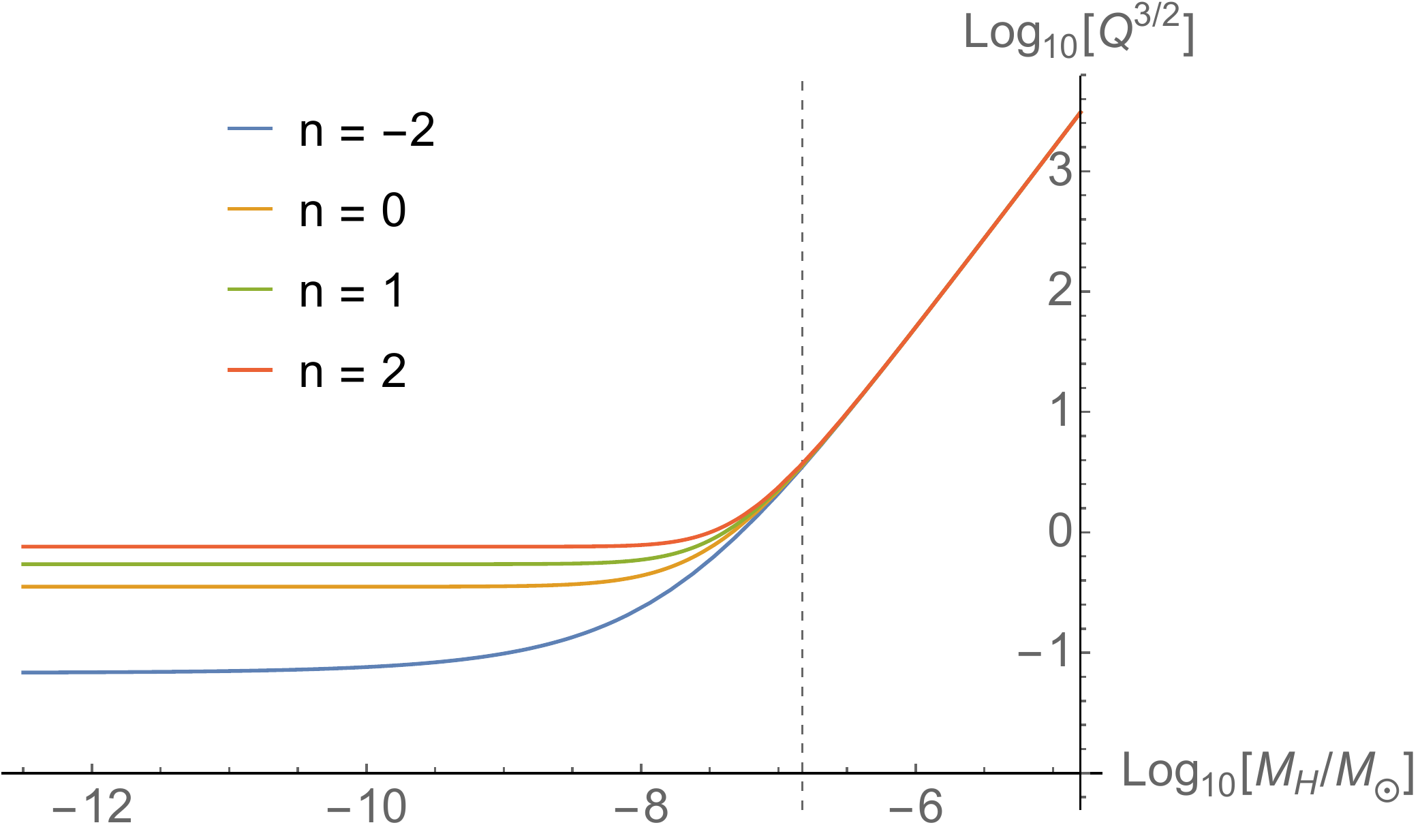}
		\hfill
		\includegraphics[width=70mm]{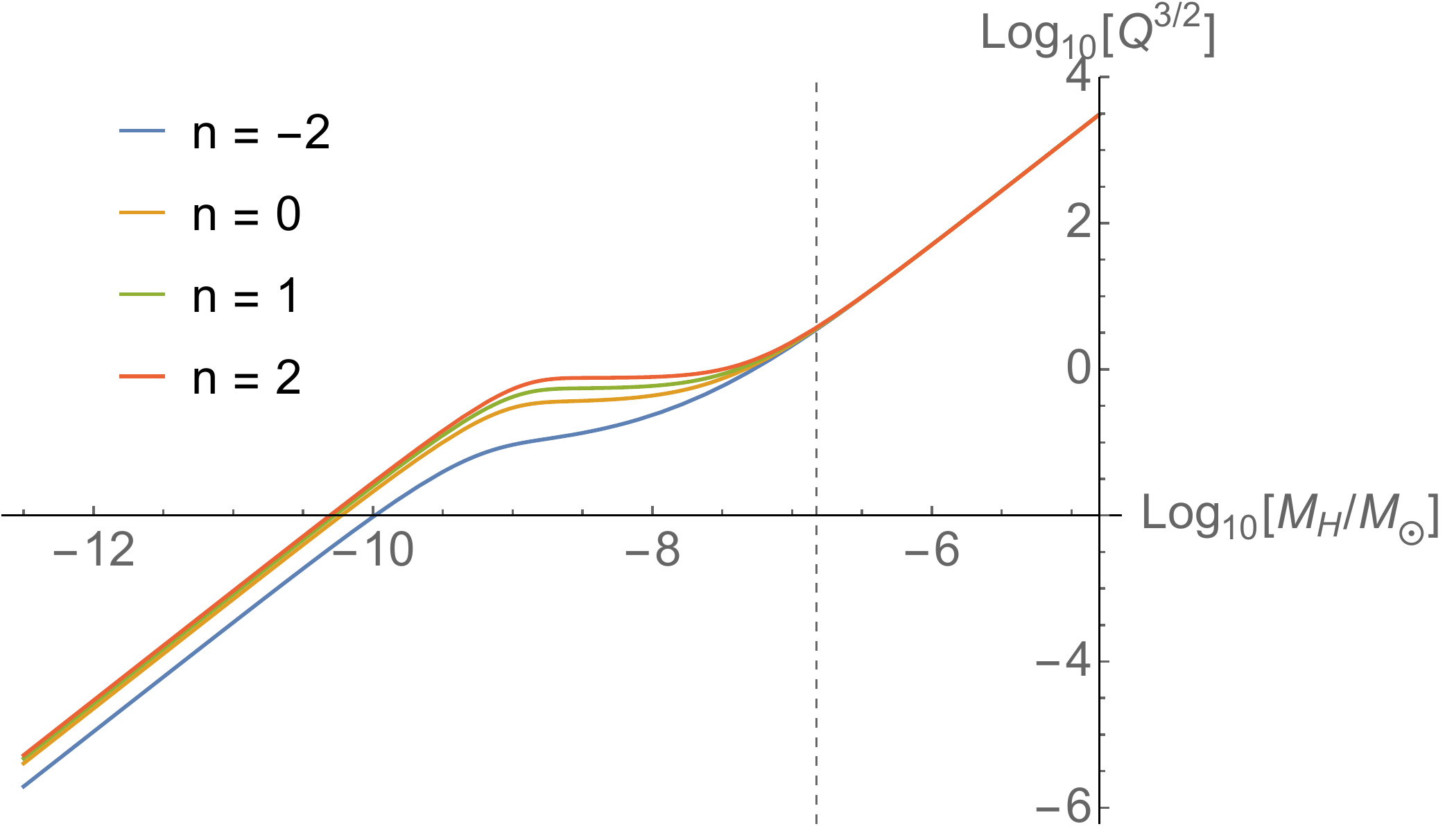}
	\end{center}
	\caption{
		\label{fig_Q}The $Q$ factor of the broken power-law spectrum (left panel) and the trapezoidal spectrum (right panel) with the pivot scale $k_0 = k_{\rm min}$ corresponding to a horizon mass $ M_{H0} = 1.5\times 10^{-7} M_\odot$ and $k_{\rm max} = 10 k_{\rm min}$ (dashed line). 
	}
\end{figure}

The spectral factor $Q$ for the trapezoidal templates takes the form of
\begin{align}
Q = \frac{1}{3} 
\frac{\Gamma\left(\frac{5+n}{2}, k_{\rm min}^2R^2\right) - \Gamma\left(\frac{5+n}{2}, k_{\rm max}^2R^2\right)}{\Gamma\left(\frac{3+n}{2}, k_{\rm min}^2R^2\right) - \Gamma\left(\frac{3+n}{2}, k_{\rm max}^2R^2\right)}.
\end{align}
As shown in Figure~\ref{fig_Q}, one can see that the behavior of $Q^{3/2}$ is the same as that of the broken power-law templates for $k_{\rm max}R > 1$. In the limit of $k_{\rm max}R \ll 1$ it shows that $Q^{3/2} \ll 1$ for all choices of $n$.

\section{Extended mass functions}\label{sec:mass_function}

So far we have assumed that the PBH mass $M$ is fixed by the Hubble horizon mass with the uniform relation $M = M_H$.
In this section we proceed one step further by taking into account the BH mass to density correlation, that is $\Delta = \Delta(M)$, induced via the effect of near-critical gravitational collapse \cite{Niemeyer:1997mt,Yokoyama:1998pt}. Due to the critical effect, the PBHs formed at each Hubble mass scale $M_H$ spans a distribution in $M$, that is $\beta_{\rm PBH} =\beta(M, M_H)\equiv d\Omega(M_H)/d\ln M$, where $\Omega(M_H)$ is the PBH fractional density at a given $M_H$. It is perhaps convenient to regard $M_H$ as the time parameter in this discussion. At the end, we sum up the contribution at each Hubble time to find the total mass distribution seen after matter-radiation equality.


\subsection{point-like peaks}
We have shown that the fractional density derived from the special peak statistics with point-like reduction agrees with the prediction from the Press-Schechter (PS) approach for the broken power-law templates with $n \leq 1$ and for the trapezoidal templates with all choices of $n$.
In this section, we use focus on the result based on the PS method.
The extended PBH mass function derived from the PS approach is a generalization of the Carr's formula \cite{Carr:1975qj} with density peaks in terms of PBH masses.
The range of the parameter relevant to our question is given by $\Delta_{\rm max} > \Delta \geq \Delta_c$, where $\Delta_c$ is the threshold density for BH formation and $\Delta_{\rm max}$ is a cutoff. 
The exact value of $\Delta_c$ may have a correlation with the input inflationary spectrum \cite{Germani:2019zez}. 
Note that the density contrast $\Delta$ defines on comoving hypersurfaces is identical to the spatial curvature at linear order so that it removes the possible background bias due to superhorizon curvature perturbations \cite{Green:2004wb,Young:2014ana}. 

For a given probability distribution $P(\Delta)$, the fraction of the density that collapses into BHs at the epoch with a horizon mass $M_H = 4\pi\rho/(3 H^3)$, where $H$ is the Hubble parameter, is led by the formalism
\begin{align}\label{density_fraction_PS}
\Omega(M_H) = \frac{1}{M_H}\int_{\Delta_c}^{\Delta_{\rm max}} P(\Delta) M(\Delta) d\Delta.
\end{align}
In realistic cases, $P(\Delta)$ is a rapidly declining function above $\Delta_c$ so that one can usually perform the replacement $\Delta_{\rm max} \rightarrow \infty$. To make a clear comparison with the peak theory (without exactly spherical symmetry), we focus on the Gaussian distribution as
\begin{align}
P_{\rm PS}(\Delta) = \frac{1}{\sqrt{2\pi} \sigma_\Delta} \exp \left[-\frac{\Delta^2}{2\sigma_\Delta^2}\right], 
\end{align}
where $\sigma_\Delta$ given by \eqref{variance_gaussian_window} is the variance of $\Delta$ that captures the information of the power spectrum $\mathcal{P}_\Delta$. 
Again, to make a clear comparison, we shall use the same choice of window function in the later discussion on peak theory. 

We may also derive the one-variable effective probability distribution function from the special peak statistics \eqref{n_spk inter1} as
\begin{align}
P_{\rm spk}(\Delta) = \frac{1}{2\pi \sigma_\Delta}\sqrt{ \frac{\pi}{2}}
\left(1 + \text{erf}\left[\frac{\gamma\Delta}{\sqrt{2(1-\gamma^2)}\sigma_\Delta}\right]\right)
\exp \left[-\frac{\Delta^2}{2\sigma_\Delta^2}\right], 
\end{align}
where we neglect the factor $2$ in this definition. In the limit of $\Delta/\sigma_\Delta \rightarrow \infty$, one finds $P_{\rm spk} \rightarrow P_{\rm PS}$. We focus on the mass functions from inflationary spectra that satisfy the high peak approximation $P_{\rm spk} = P_{\rm PS}$. 

If PBHs are formed exactly with the horizon mass, namely $M = M_H$, then \eqref{density_fraction_PS} reproduces the previous results \cite{Young:2014ana} (upto a factor of $2$). However, the effect of critical collapse shows that the PBH masses should have a distribution near $M_H$ \cite{Niemeyer:1997mt}, which is often parametrized via the scaling formula as
\begin{align}\label{critical_collapse_definition}
M = K M_H \left( \Delta - \Delta_c\right)^{\gamma_m}.
\end{align}
Here $K = 3.3$ and $\gamma_m = 0.35$ are numerical constants. The profile dependence of $K$ and $\Delta_c$ \cite{Yoo:2018esr,Kalaja:2019uju}, if considered, should be applied to both statistical methods. This simple extension allows us to rewrite the density contrast in terms of the PBH mass as $\Delta = \Delta(M)$. The differential PBH density $\beta(M,M_H) \equiv d \Omega(M_H)/d\ln M$ at $M_H$ according to \eqref{density_fraction_PS} is therefore obtained as
\begin{align}\label{diff_mass_function_PS}
\beta_{\rm PS} = \frac{K}{\sqrt{2\pi}\gamma_m \sigma_\Delta}\left(\frac{M}{K M_H}\right)^{1+1/\gamma_m} \exp\left[-\frac{\Delta^2(M)}{2\sigma_\Delta^2}\right],
\end{align}
where $\Delta(M) = (M/(K M_H))^{1/\gamma_m}+\Delta_c$. 

Having in mind that PBHs behave as matter in the radiation dominated universe, the relative density $\rho_{\rm PBH}/\rho \sim a$ is growing with time. By using the approximation $w = 1/3$ as a constant until matter-raidation equality \cite{Byrnes:2018clq,Wang:2019kaf}, the mass function at $a = a_{eq}$ reads $\beta_{eq}(M, M_H) = (a_{eq}/a)\beta(M, M_H) = (M_{Heq}/M_H)^{1/2}\beta(M, M_H)$. Finally, we arrive at the total mass distribution for PBHs formed during the radiation dominated epoches by the integration over $M_H$ as
\begin{align}\label{total_mass_function_PS}
f_{\rm PS}(M) = \frac{1}{\Omega_{\rm DM}} \int^{\ln M_{Heq}}_{\ln M_{\rm min}} \left(\frac{M_{Heq}}{M_H}\right)^{1/2} \beta_{\rm PS} d \ln M_H.
\end{align}
We remark that the lower limit $M_{\rm min}$ comes from the upper bound $\Delta_{\rm max}$ for the density perturbation. Applying a conservative condition $\Delta_{\rm max} = 2\Delta_c$ for the validity of the formula \eqref{critical_collapse_definition}, we find that $M_{\rm min} = M/(K \Delta_c^{\gamma_m})$.

\subsection{general peaks}


We now compute the extended mass function from peak statistics \cite{Bardeen:1985tr} without imposing spherical symmetry to the density perturbation.
One can express peaks in terms of BH and horizon masses via \eqref{critical_collapse_definition} as
\begin{align}\label{critical_collapse_peak}
\nu &= \frac{1}{\sigma_\Delta} \left(\frac{M}{K M_H}\right)^{1/\gamma_m} +\nu_c, \\
d\nu &=  \frac{1}{\gamma_m\sigma_\Delta} \left(\frac{M}{K M_H}\right)^{1/\gamma_m} d\ln M.
\end{align}
Here $\nu_c = \Delta_c/\sigma_\Delta$ and $\sigma_\Delta (R)$ can be obtained in terms of $M_H$ through \eqref{variance_gaussian_window} and \eqref{horizon_mass to smoothing}.
The high peak expansion of $N_{BBKS}$ \eqref{N_BBKS} in the limit of $\nu \gg 1$ is useful when we are only interest in inflationary spectra of the narrow spike shape. Following the findings in the previous section, the differential number density can be reduced as
\begin{align}\label{N_BBKS_highpeak}
N_{\rm BBKS}(\nu) d\nu \approx \frac{Q^{3/2}}{(2\pi)^2} 
\left(\nu^3 -3\frac{\nu}{\gamma^2}\right)e^{-\nu^2/2}d\nu,
\end{align}
where $Q^{3/2} = \gamma^3R_\ast^{-3}$ and the high-peak approximation is valid if $\nu^3 > 3\nu/\gamma^2$. Our numerical results indicate that \eqref{N_BBKS_highpeak} is a good approximation for the broken power-law templates with $n \lesssim 1$ and for the trapezoidal templates with $n \lesssim 1.2$. 

If the BH mass is just coincides with $M_H$, the fractional density of peaks that satisfy the criterion of PBH formation at the epoch with a fixed horizon mass $M_H$ is approximated by $\Omega_{M_H}\approx n_{\rm pk}(\nu, M)M/\rho_{M_H}$ \cite{Green:2004wb,Young:2014ana}. With the extended correlation $\nu = \nu (M)$ led by the critical collapse \eqref{critical_collapse_peak}, the fractional density of PBH is now written as
\begin{align}
\Omega_{\rm PBH}(M_H) &= \frac{1}{\rho(M_H)} \int_{\nu_c}^{\nu_{\rm max}} \rho_{\rm PBH}(\nu) d\nu, \\
&=  \frac{V(R)}{M_H} \int_{\nu_c}^{\nu_{\rm max}} N_{\rm BBKS}(\nu) M(\nu) d\nu,
\end{align}
where $\nu_{\rm max} = \Delta_{\rm max}/\sigma_\Delta$ and we have fixed the smoothing scale $R$ with the comoving horizon. $ V(R)= (\sqrt{2\pi}R)^3$ is the volume of the Gaussian window function that satisfies the normalization condition $V(R)^{-1}\int W(x, R)d^3x =1$ with $W(x, R) = \exp[-x^2/(2R^2)]$ in the real space. Again, the derivative of $\Omega(M_H)$ with respect to the logarithmic of $M$ gives the differential PBH density as
\begin{align}\label{diff_mass_function_pk}
\beta_{\rm BBKS} = \frac{K}{\sqrt{2\pi}\gamma_m \sigma_\Delta} Q^{3/2}
\left(\frac{M}{K M_H}\right)^{1+1/\gamma_m} 
\left(\nu^3 - 3\nu\right)e^{-\nu^2/2},
\end{align}
where the factor $Q = Q(M_H) = \sigma_1^2 R^2/(3\sigma_\Delta^2)$.
The total mass distribution accounted for PBHs formed before matter-radiation equality is computed by the same formula as \eqref{total_mass_function_PS}, which reads
\begin{align}\label{total_mass_function_peak}
f_{\rm BBKS}(M) = \frac{1}{\Omega_{\rm DM}} \int^{\ln M_{Heq}}_{\ln M_{\rm min}} \left(\frac{M_{Heq}}{M_H}\right)^{1/2} \beta_{\rm BBKS} d \ln M_H.
\end{align}
Comparing \eqref{diff_mass_function_PS} with \eqref{diff_mass_function_pk} one can see the difference $\beta_{\rm BBKS}/\beta_{\rm PS} \approx Q^{3/2}\nu^3$ for $\nu \gg 1$. In general, $\nu^3$ enhances the amplitude of the BBKS mass function and the $Q$ factorizes the spatial dependence of the input inflationary spectrum.



\subsection{systematic bias or underestimation?}

\begin{figure}
	\begin{center}
		\includegraphics[width=65mm]{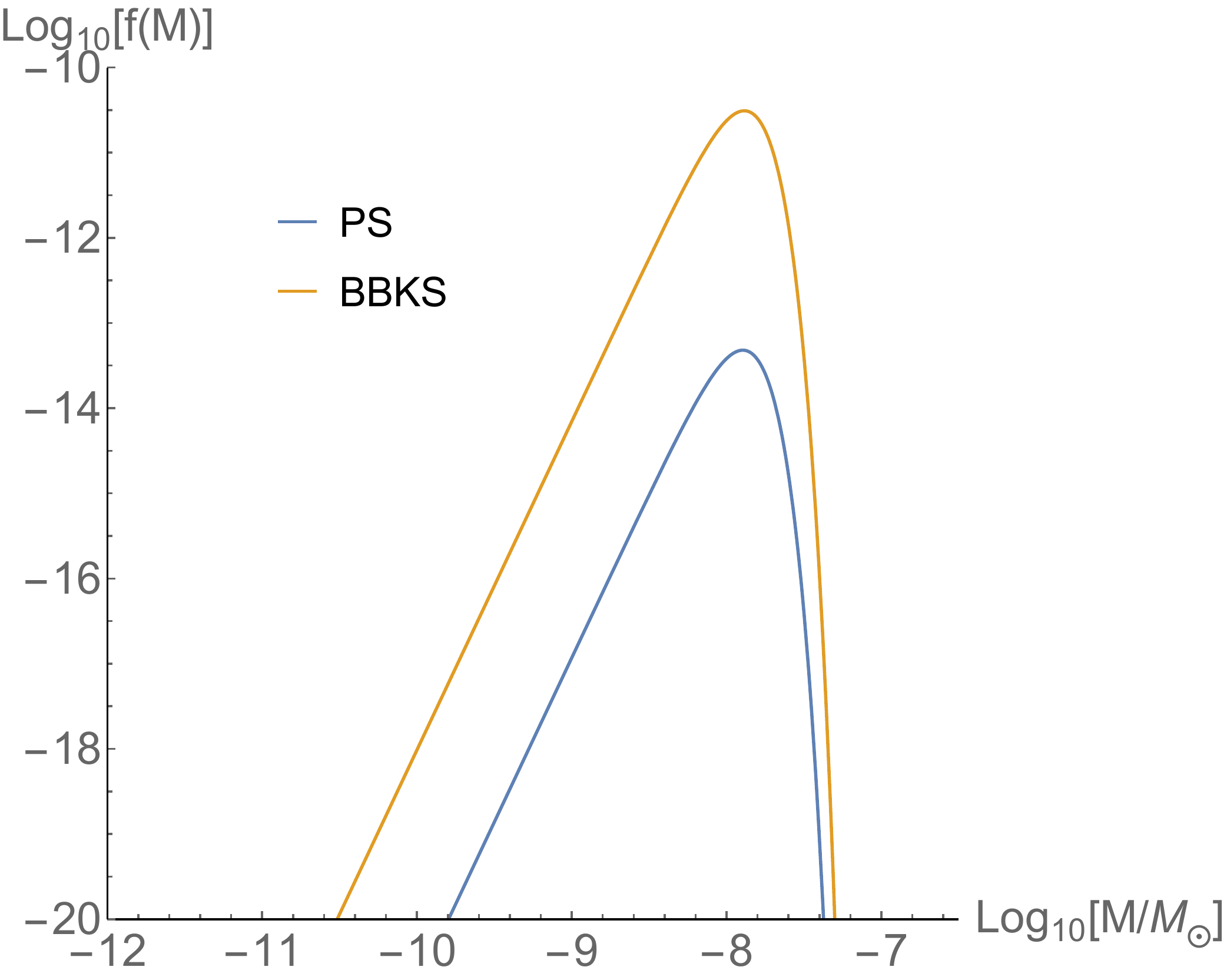}
		\hfill
		\includegraphics[width=65mm]{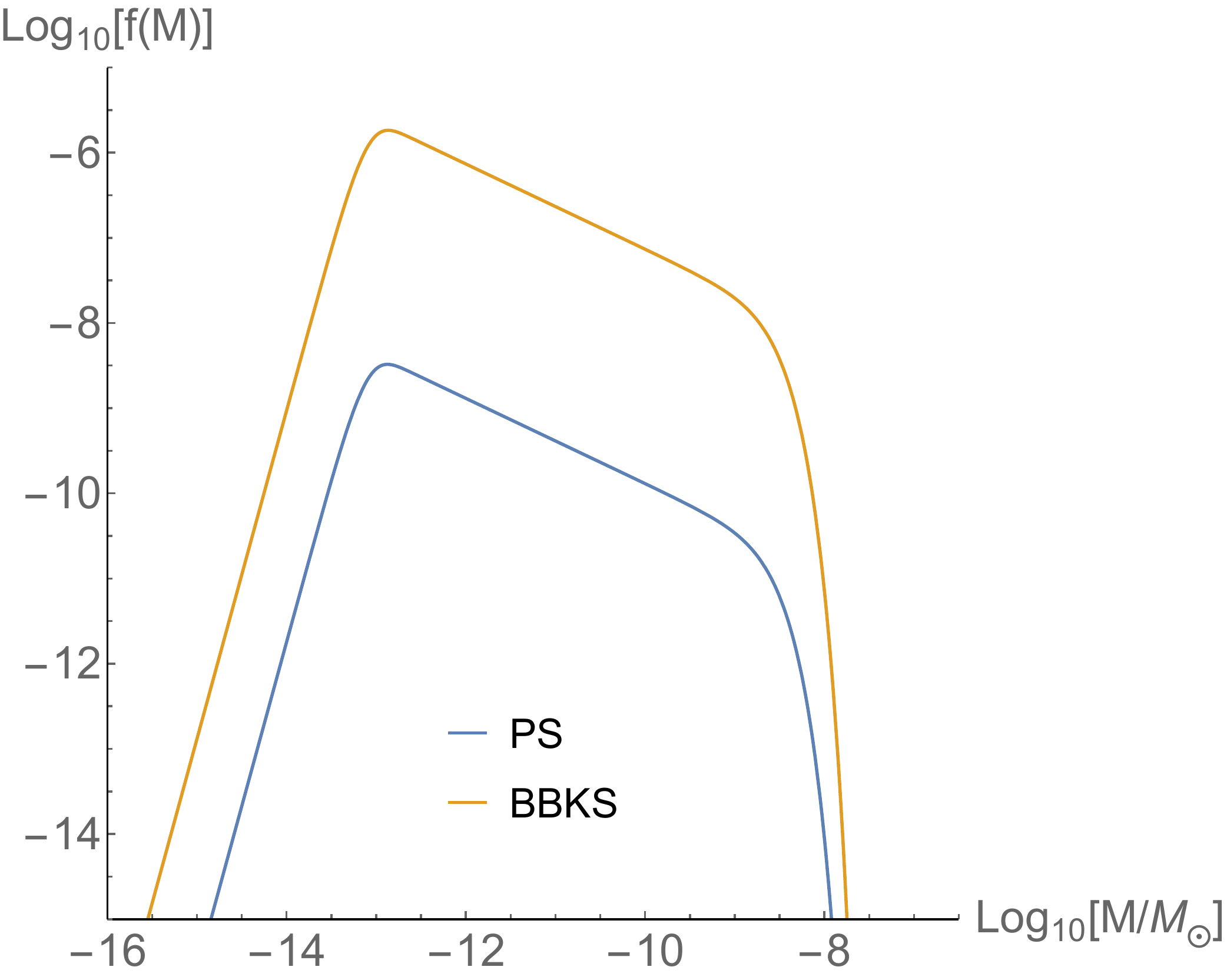}
	\end{center}
	\caption{
		\label{fig_f_1000}The mass function $f(M)$ from the trapezoidal templates with the pivot scale $k_{\rm min}$ corresponding to a horizon mass $ M_{H0} = 1.5\times 10^{-7} M_\odot$ and $k_{\rm max} = 1000 k_{\rm min}$. The spectral indices are chosen as $n = -1$ with $A_\zeta = 0.05$  (left panel) and $n = 1$ with $A_\zeta = 0.02$ (right panel). 
	}
\end{figure}

\begin{figure}
	\begin{center}
		\includegraphics[width=75mm]{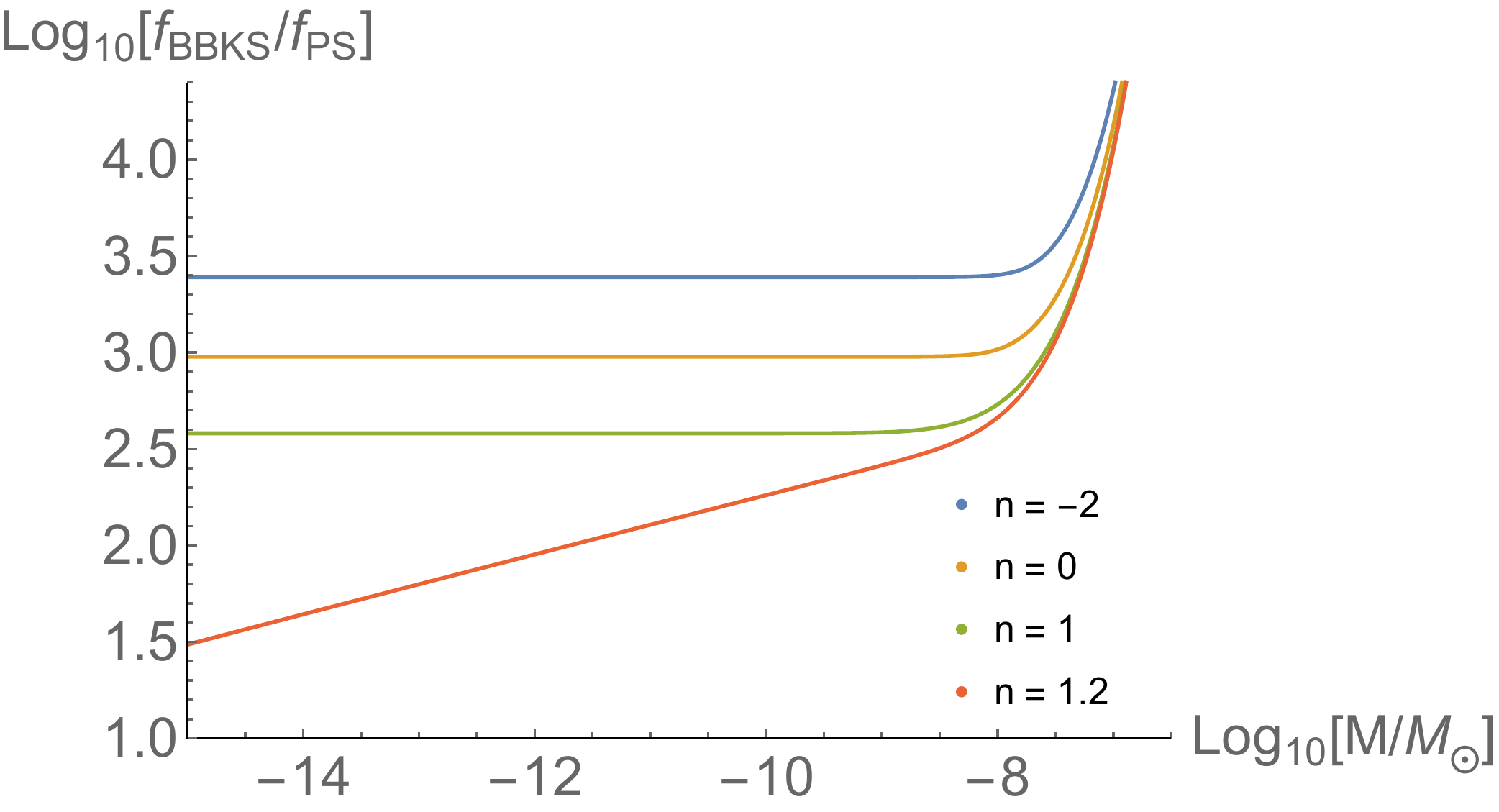}
		\hfill
		\includegraphics[width=75mm]{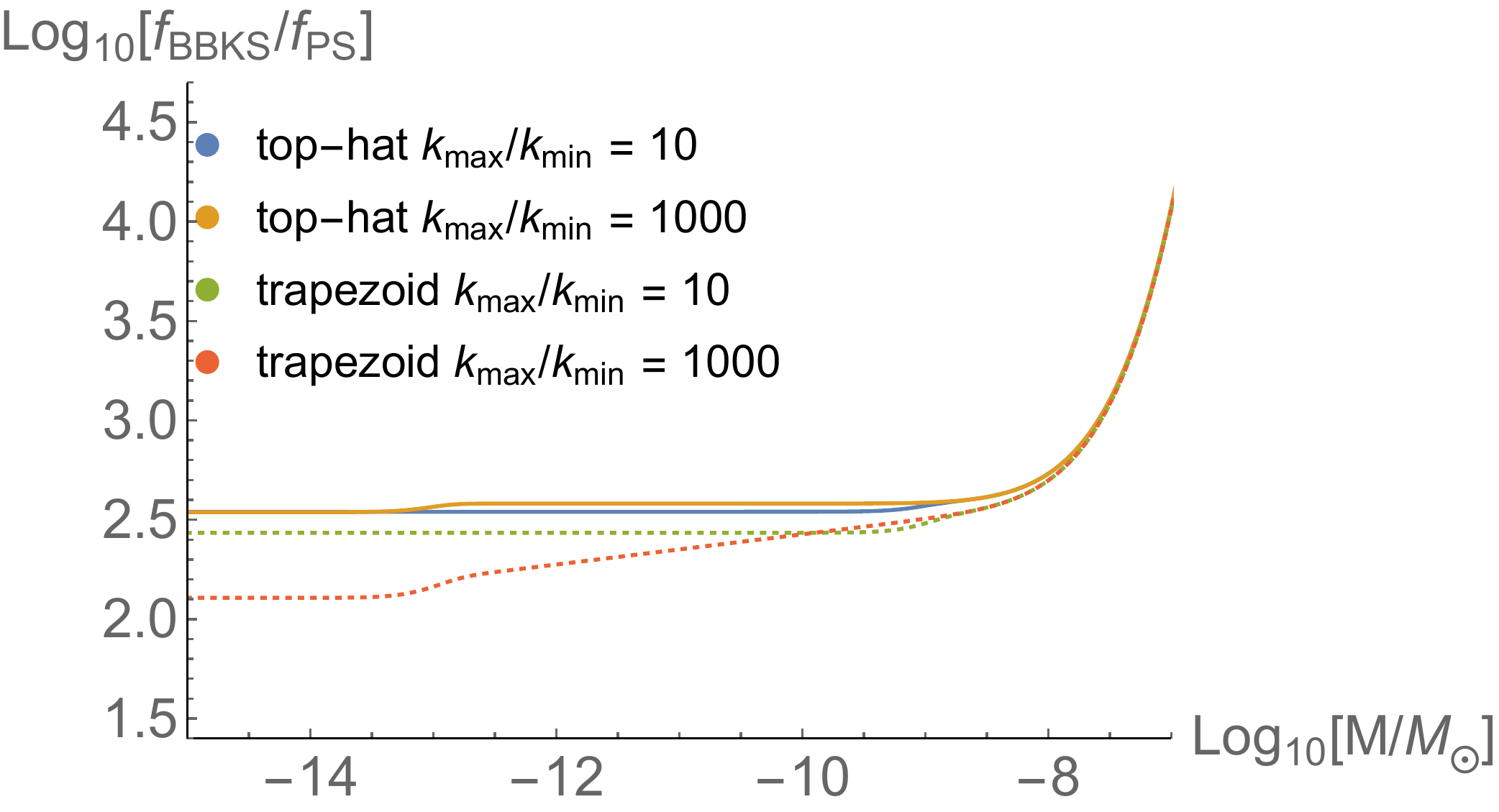}
	\end{center}
	\caption{
		\label{fig_f_ratio}The ratio of the mass functions based on the broken power-law spectrum (left panel) and the trapezoidal templates (right panel) with the pivot scale $k_0 = k_{\rm min}$ corresponding to a horizon mass $ M_{H0} = 1.5\times 10^{-7} M_\odot$ for various $n$. In right panel, $n = 1$ is used for the top-hat spectrum and $n = 1.1$ (dotted lines) is used for the trapezoidal spectrum.
	}
\end{figure}

	For a spectral index $n \lesssim 1$, the equivalence $\beta_{\rm PS} = \beta_{\rm spk}$ shown in the previous section indicates that the PS method technically treats the high-peak density perturbations as unphysical point-like objects with zero spatial dimension. In contrast with the point process in the BBKS method \eqref{n_BBKS point process}, the determinant $\text{det} \vert \mathbf{z}\vert$ carring the information of the 3-dimensional configuration is encoded in the function $G$ in \eqref{G_general} after the reduction of statistical variables. This is the main origin of the analytic relation $\beta_{\rm BBKS}/\beta_{\rm PS} \sim \nu_c^3$ in the high-peak limit. Since realistic density peaks are 3-dimensional in space, the systematic bias $\nu_c^3\sim \mathcal{O}(10)$ found in \cite{Young:2014ana} (based on slightly blue-tilted spectra) in fact implies a factor of underestimation for the PBH abundance estimated via the PS method.
	Having shown that for non blue-tilted spectra with $n \leq 1$ the factor $\nu_c^3 $ can be $ \mathcal{O}(10^2)$ or larger,
in this section we compare the mass functions result from the conventional PS method \eqref{total_mass_function_PS} (or the point-like peak statistics with spherical symmetry) and the BBKS method \eqref{total_mass_function_peak} (the general 3-dimensional peak statistics). The systematic bias between $f_{\rm BBKS}$ and $f_{\rm PS}$ is summarized in Figure~\ref{fig_f_ratio} for both the broken power-law and the trapezoidal templates (see Figure~\ref{fig_f_1000}).

For $n < 1$ the mass function $f(M)$ exhibits a spike slightly lower than the pivot horizon mass $M_{H0}$ (corresponding to $k_0$ for the broken power-law templates and $k_{\rm min}$ for the trapezoidal templates), where we refer the shape in this parameter space as the spiky mass spectrum.
The spike scale $M_{\rm spike}$ such that $f(M_{\rm spike})$ is maximum can be found numerically, and it shows that $M_{\rm spike} \approx 10^{-7.9} M_\odot$ with $n = -1$ for both templates. The scale $M_{\rm spike}$ increases towards $M_{H0}$ with the decrease of the spectral index $n$.

The ratio $f_{\rm BBKS}/f_{\rm PS}$ for $n < 1$ is larger than $10^{2.5}$. This large discrepancy from input spectra in the narrow-spike shape was not recognized by previous studies. In the limit of $M \rightarrow 0$ (namely $\nu \rightarrow \nu_c$), one finds
\begin{align}
\frac{f_{\rm BBKS}}{f_{\rm PS}} \approx \left.\frac{\beta_{\rm BBKS}}{\beta_{\rm PS}}\right\vert_{M_H = M_{\rm spike}} \simeq Q^{3/2}\nu_c^3,
\end{align}
where $Q$ and $\nu_c$ are evaluated at $M_H = M_{\rm spike}$. For $n = -2$ we find $f_{\rm BBKS}/f_{\rm PS} \sim 10^{3.4}$ in the limit of $M \ll M_{H0}$.
We remark that for both templates in the limit of $M \gg M_H$ the BH formation rate is too rare so that the ratio $f_{\rm BBKS}/f_{\rm PS}$ is very sensitive to a small change in the mass parameter $M$. The sharply enhanced ratio in the $M > M_{H0}$ limit shall not have an important meaning since $f(M)$ is rapidly dropped off due to the effect of critical collapse. 

For $n = 1$ we observe $f_{\rm BBKS}/f_{\rm PS} \simeq 10^{2.5}$ for both broken power-law and trapezoidal templates. This is the divide for the red and blue tilted spectrum and increasing the broadness of the top-hat spectrum does not change the ratio $f_{\rm BBKS}/f_{\rm PS}$ much. Since $Q \approx 1$ in the plateau region of $\mathcal{P}_\zeta(k)$ with the step or the top-hat shape (see Figure~\ref{fig_Q}), the bias $f_{\rm BBKS}/f_{\rm PS} \simeq \nu_c^3$ is led by the peak value $\nu_c \approx 8.8$.
Note that for the top-hat spectrum the number density of PBHs formed between $k_{\rm min}^{-1}$ and $k_{\rm max}^{-1}$ is the same, as shown in Figure~\ref{fig_beta_tra1}. Therefore the BH mass corresponding to $k_{\rm max}$ has the largest weight in the mass function due to the relative growth of the PBH density in radiation domination (see also \cite{DeLuca:2020ioi}).

For templates with $n > 1$ the high peak approximation \eqref{N_BBKS_highpeak} in general breaks down. However by using indices slightly larger than $n = 1$ we observe interesting tendency for blue-tilted spectra. In Figure~\ref{fig_f_ratio} we find that the ratio $f_{\rm BBKS}/f_{\rm PS}$ can be smaller than $10^{2.5}$ with slightly blue templates, which recovers the results of \cite{Young:2014ana}. 
For the broken power-law template with $n = 1.2$, the abundance is dominated by $M_H$ in the small mass limit so that
\begin{align}
\frac{f_{\rm pk}}{f_{\rm PS}} \approx \left.\frac{\beta_{\rm pk}}{\beta_{\rm PS}}\right\vert_{M_H = M_{\rm min}} \simeq \nu_c^3,
\end{align}
where $\nu_c$ is computed at $M_H = M_{\rm min}$ and $M_{\rm min}$ is our lower bound of the numerical computation. 
The decrease of the ratio $f_{\rm BBKS}/f_{\rm PS}$ is due to the enhance of $\sigma_\Delta$ in the limit of $M_H \rightarrow 0$.



\section{Summary}
The selecting process that PBHs only form at local maxima of the density perturbation invokes a construction of joint probability distribution for the random field $\Delta$ with its first and second spatial derivatives.
	We have shown that, $n_{\rm spk}\equiv \beta_{\rm spk}$, the number density of PBHs evaluated by the ensemble average of dimensionless point-like peaks \eqref{n_spk full result} coincides with, $\beta_{\rm PS}$, the standard PBH abundance by virtue of the Press-Schechter method \eqref{n_PS}. 
	The number density $n_{\rm spk}$, however, is a collection of unphysical density peaks with zero dimension in space, and is introduced to clarify the inaccuracy made by the PS method.
	The discrepancy of the PBH abundance from the two approaches, $\beta_{\rm PS}$ and $ \beta_{\rm spk}$, is negligible unless using a super blue-tilted inflationary spectrum. 

The standard BBKS peak statistics \cite{Bardeen:1985tr} uses conditional point process in 3-dimensional space and in general allows a finite deviation from spherical symmetry. 
	By using the high-peak expansion in the limit of $\nu \gg 1/\gamma$, we reproduced the analytic relation $\beta_{\rm BBKS}/\beta_{\rm PS} \approx \beta_{\rm BBKS}/\beta_{\rm spk} \sim \nu_c^3$, where $\beta_{\rm BBKS}/\beta_{\rm PS} \sim \nu_c^3$ was firstly recognized in \cite{Young:2014ana} as a ``systematic bias'' based on blue-tilted inflationary spectra. With numerical examinations, we have confirmed that the high-peak expansion \eqref{G_highpeak} in general breaks down for blue-tilted spectra, yet $\beta_{\rm BBKS}/\beta_{\rm spk} \approx Q^{3/2}\nu_c^3 \sim 10^2$ is a good analytic estimation for inflationary spectra in the flat or red-tilted (spiky) shape. Given that realistic density peaks which would form PBHs are 3-dimensional in space, the equivalence $\beta_{\rm PS} = \beta_{\rm spk} \equiv n_{\rm spk}$ at leading order in the high peak limit implies an underestimation (rather than a pure systematic bias) for the PBH abundance via the Press-Schechter approach.

We have computed the extended mass function, $f(M)$, for BH formation under the effect of critical collapse. 
For the first time, a generic discrepancy $f_{\rm BBKS}/f_{\rm PS} \gtrsim 10^{2.5}$ in all mass range has been reported, and for the inflationary spectrum in the narrow-spike class (the favorable shape for realizing PBHs as all dark matter) the systematic difference can be raised to $\sim 10^{3.4}$. 
Note that $f_{\rm PS}$ also stands for the prediction for the point-like peak statistics, and
the discrepancy led by $\nu_c^3$ (namely the factor of underestimated PBH abundance) is significantly larger than the findings based on blue-tilted spectra \cite{Young:2014ana}.

We remark that
	the point-like peak theory introduced in this work provides a clear physical interpretation for the factor $\nu_c^3 \sim \beta_{\rm BBKS}/\beta_{\rm PS}$, as
 the ratio $n_{\rm BBKS}/n_{\rm spk} $ compares number density of peaks reside in
  the 3-dimensional space to unphysical peaks with zero dimension.
In a more formal calculation of the PBH abundance, the subhorizon dynamics of $\Delta$ or the non-linear effect of $\zeta$ to $\Delta$ shall not affect the enhancement due to the volume effect, as long as PBH formation is only valid for high sigma peaks $\nu_c \gg 1$. 
We expect a same conclusion by changing the choices of smoothing window function. 
	While all presented methods for computing the PBH mass function from inflation can fit the observational constraints with different model parameters, our results provide a theoretical justification that using the PS method may at least miss a physical effect led by the real dimensionality of the high peak density perturbations.
The largely enhanced mass function due to the volume effect in a general 3-dimensional configuration would imply a more stringent constraint on the spectral amplitude from inflation \cite{Carr:2017jsz,Green:2016xgy,Kuhnel:2017pwq}, especially for the spectrum in the topic of PBH as all dark matter, if based on the estimation via the PS method. A similar conclusion might also have impact on the topic of PBH dark matter for the future space-based observations \cite{Cai:2018dig,Bartolo:2018evs}.



\acknowledgments
The authors thank Christian Byrnes, Cristiano Germani, Minxi He, Misao Sasaki and Pi Shi for the helpful comments and discussions.
The author thanks Jun'ichi Yokoyama for the initiative idea of this project.
We acknowledge the workshop ``Focus week on primordial black holes'' at Kavli IPMU. 
Y.-P. Wu was supported by JSPS International Research Fellows and JSPS KAKENHI Grant-in-Aid for Scientific Research No. 17F17322,
 and is supported by the the ANR ACHN 2015 grant (“TheIntricateDark” project).



\end{document}